\documentclass[twocolumn]{article}
\usepackage[T1]{fontenc}
\usepackage[utf8]{inputenc}
\usepackage[a4paper]{geometry}
\usepackage[english]{babel}
\usepackage{babelbib}
\usepackage{textcomp}
\usepackage[unicode=true]{hyperref}
\hypersetup{colorlinks=true,citecolor=blue,urlcolor=blue,linkcolor=blue}
 \makeatletter

\usepackage{graphicx}
\usepackage{siunitx}
\usepackage{amsmath}
\usepackage[font=small,labelfont=bf]{caption}
\usepackage{subcaption}
\usepackage{url}
\usepackage[modulo,switch]{lineno}
\modulolinenumbers[1]

\@ifundefined{date}{}{\date{}}
\makeatother

\begin{document}

\title{\textbf{Denoising of photogrammetric dummy head ear point clouds for individual Head-Related Transfer Functions computation}}

\author{Fabio Di Giusto$^{1,2,}\footnote{Corresponding author: \href{mailto:fabio.digiusto@kuleuven.be}{fabio.digiusto@kuleuven.be}}$,
Francesc Lluís$^{3}$,
Sjoerd van Ophem$^{1,2,4}$,
Elke Deckers$^{5,2}$\\
\footnotesize $^1$ Department of Mechanical Engineering, KU Leuven, B-3001, Leuven, Belgium\\
\footnotesize $^2$ Flanders Make@KU Leuven, Belgium\\
\footnotesize $^3$ Bang \& Olufsen A/S, 7600, Struer, Denmark\\
\footnotesize $^4$ Institute of Sound and Vibration Research, University of Southampton, SO17 1BJ, Southampton, United Kingdom\\
\footnotesize $^5$ Department of Mechanical Engineering, KU Leuven, Campus Diepenbeek, 3590, Diepenbeek, Belgium\\
}

\maketitle\thispagestyle{empty}

\begin{abstract}
Individual Head-Related Transfer Functions (HRTFs), crucial for realistic virtual audio rendering, can be efficiently numerically computed from precise three-dimensional head and ear scans. While photogrammetry scanning is promising, it generally lacks accuracy, leading to HRTFs showing significant perceptual deviation from reference data, mainly due to scanning errors affecting the most occluded pinna structures. This paper examines the application of Deep Neural Networks (DNNs) for denoising photogrammetric ear scans. Several DNNs, fine-tuned on pinna samples corrupted with synthetic error modelled to mimic that observed in photogrammetric dummy head scans, are tested and benchmarked against a classical denoising method. One DNN is further modified and retrained to enhance its denoising performance. The comparison of HRTFs derived from original and denoised scans against reference data shows that the best-performing DNN marginally reduces the deviation of photogrammetric dummy head HRTFs to levels closer to accurately measured ones. Additionally, correlation analysis between geometric and HRTF metrics, computed on the scanned point clouds and their corresponding HRTFs, is used to identify key measures for evaluating the deviation between target and reference scans. These findings are expected to guide the selection of relevant loss functions and foster improvements in this and similar DNN models.
\end{abstract}

\section{Introduction}
\label{sec:1}
Binaural audio rendering describes the spatialisation of sound by mimicking the cues used by the human auditory system to locate sound sources. These localisation cues entail binaural cues, i.e.\ the Interaural Time Difference (ITD) and Interaural Level Difference (ILD), as well as monaural spectral cues \cite{Katz2018BinauralReproduction}. The first are strongly related to the size and shape of a listener's head, while the latter depend primarily on pinna morphology and are considered to play a crucial role in elevation perception and front-back discrimination, especially when the binaural cues are ambiguous \cite{Baumgartner2014ModelingListeners}. All these spatial cues are contained in the Head-Related Transfer Functions (HRTFs), describing the spectro-temporal filtering effect generated by the interaction of a sound field with a listener's ears, head, and torso morphology. Thus, by using headphones to present a sound filtered with individual HRTFs, it is possible to obtain a reliable spatialisation of a virtual sound source \cite{Pollack2022PerspectiveOverview}. However, these filters are highly personal, given the distinctive ear anatomy of each listener, and individualised HRTFs are considered necessary for the most accurate audio rendering even in immersive Virtual Reality (VR) scenes, although additional localisation cues can be leveraged with this approach, i.e.\ dynamic and visual cues \cite{Jenny2020UsabilityLocalization}. Thus, a challenge in the acquisition of HRTFs is that it should ideally be done on an individual basis. 

Individual HRTFs are mainly obtained through: (i) experimental measurements, (ii) numerical calculations or (iii) personalisation techniques. The first method typically relies on free-field measurements of impulse responses between one or more loudspeakers and two microphones placed in a listener's ear canals. The second approach is based on estimating the pressure generated by multiple sound sources at the ear canals of a 3D-scanned individual geometry, using numerical techniques such as the Boundary Element Method (BEM) or Finite Element Method (FEM). The personalisation techniques generally work by selecting or adapting non-individual HRTFs to a specific listener using objective metrics, e.g.\ anthropometric measures, or subjective data, e.g.\ spatial sound perception with different HRTFs \cite{Pollack2022PerspectiveOverview}. While HRTF personalisation techniques are simple and cost-effective, their application yields moderate results, and their accuracy is inferior to measured or numerically computed HRTFs \cite{Zhong2014Head-RelatedDisplay}. The current study mainly focuses on the numerical computation of HRTFs, given the accuracy and potential scalability of this method owing to the advances in affordable scanning techniques. However, the main challenge of this approach lies in the scanning of a subject's ear geometry, since accurate HRTF computation requires a sufficiently high precision, around \SI{1}{\milli\meter}, at the pinnae \cite{Ziegelwanger2013CalculationQuality}. Several techniques can be used for acquiring the geometry, e.g.\ Magnetic Resonance Imaging, Computed Tomography, laser or structured-light scanning. These methods tend to yield results considered accurate enough for HRTF computation but have several downsides related to expensive equipment and exposure to strong magnetic fields or high radiation, which might require using a cast of a listener's ear geometry \cite{Pollack2022PerspectiveOverview}. An alternative and affordable scanning technique is photogrammetry, which has the advantage of being easily scalable as it only requires widely available camera sensors, e.g.\ smartphone or digital cameras. Nonetheless, the drawback of this technique is its low accuracy due to the inherent scanning error, especially affecting ear locations with limited visibility due to self occlusion of the complex pinna geometry, such as the pinna cavities. The scanning error tends to affect the spectral peaks and notches of the estimated HRTFs, thus hindering the localisation cues used for elevation perception. Moreover, extensive post-processing of the raw photogrammetric point clouds is needed to obtain suitable meshes for HRTF computation. The HRTFs computed on the photogrammetric scan of an optically treated dummy head show ITDs and ILDs close to reference measured and simulated data. Yet, the modelled sagittal plane localisation error with the acquired HRTFs is similar to that obtained with non-individual filters. This deviation is attributed to the geometric error between reference and photogrammetric scan, reaching maximum values around \SI{3}{\milli\meter} at the cymba conchae, impacting the amplitude and centre frequency of the HRTF spectral features \cite{DiGiusto2023AnalysisHead}. Furthermore, applying photogrammetry to dummy heads or ear plaster casts is considered to yield better results in comparison to scans of human subjects \cite{Reichinger2013EvaluationPinnas}, thus likely leading to even greater deviations in the related HRTFs. Therefore, additional post-processing methods to rectify the photogrammetric outcome, e.g.\ denoising techniques, or alternative ways to acquire an accurate listener's geometry, e.g.\ parametric pinna models, are thought to be necessary for the acquisition of accurate individual HRTFs \cite{DiGiusto2023AnalysisHead}.

The task of point cloud denoising has received attention in recent years, particularly leveraging Deep Learning (DL) techniques tailored for this purpose. Classical denoising methods, not based on DL, have also been developed, generally relying on optimisation approaches using geometric priors. However, these are often outperformed by DL methods employing Deep Neural Network (DNN) architectures, trained end-to-end on noisy-clean data pairs. Several DL denoising approaches have been proposed, using various architectures and techniques to reduce the real or modelled scanning errors affecting different point clouds \cite{Luo2021Score-BasedDenoising}. These methods are usually trained and tested on datasets composed of generic shapes artificially corrupted with synthetic noise, often modelled as unstructured Gaussian noise. Nonetheless, they are shown to also be effective in reducing other noise types, e.g.\ synthetic depth camera noise. However, as supervised learning methods, these approaches are unlikely to succeed on noise with characteristics that differ significantly from those encountered in the training data \cite{Rakotosaona2019PointCleanNet:Clouds}. It is worth mentioning that DL techniques could also be applied to directly synthesise or select personalised HRTFs from individual data, e.g.\ anthropometric measurements, pictures or scans of the subjects' pinnae, or perceptual feedback. Although promising, no outstanding results are yet seen with these approaches. Some limitations could stem from the fact that it is still unclear what the best parameters for HRTF individualisation are. Moreover, the amount of available training data is limited, which could hinder the generalisability of the DL models \cite{Miccini2020HRTFLearning}.

The goal of this paper is to analyse the performance of DNN-based denoising approaches on photogrammetric scans of dummy head ears, and their potential in improving the scanning outcome to obtain more accurate individual HRTFs. Three DL approaches are tested and benchmarked against each other, as well as against a classical denoising method, using several geometric metrics. To optimise the DL denoising performance for photogrammetric ear scans, a dataset of ear point clouds corrupted with synthetic photogrammetric error, based on experimentally acquired data, is created and used for the supervised training of the DNNs. The results of different model trainings are tested, ranging from the pretrained versions, trained on the original datasets of general shapes containing unstructured Gaussian noise, to models retrained from scratch or refined through fine-tuning on the ear dataset. Modifications are introduced to one DNN architecture to further improve its performance on this specific task. HRTF simulations are conducted on meshes derived from the original and denoised scans, and the results are evaluated in terms of objective and modelled perceptual HRTF metrics. Correlation analysis between geometric and HRTF metrics is used to further enhance the understanding of the link between pinna morphology and related HRTFs, and provide insights in the most effective geometric measures that can be used for HRTF individualisation approaches. Notably, these findings can provide a guideline in the selection of relevant metrics and loss functions for training personalisation methods that rely on obtaining HRTFs from individual scans. The choice of using dummy heads in this study stems from the fact that a reference geometry is generally available or easier to acquire than for human subjects \cite{Reichinger2013EvaluationPinnas}. Additionally, the availability of measured dummy head HRTF databases allows for further comparison between the numerically and experimentally acquired data \cite{Andreopoulou2015Inter-LaboratoryComparison}.

This paper is structured as follows. In Sec.~\ref{sec:2} the employed methods are outlined, including the geometric metrics used to evaluate the scanned and denoised ear point clouds in Sec.~\ref{sec:2A}, the definition of the synthetic photogrammetric scanning error and the creation of the ear dataset in Sec.~\ref{sec:2B}, the denoising models and their modifications tailored to the specific noise type under analysis in Sec.~\ref{sec:2C}, and the HRTF computation and assessment metrics in Sec.~\ref{sec:2D}. Sec.~\ref{sec:3} presents the results, detailing the outcomes of the various approaches in the denoising of ear point clouds in Sec.~\ref{sec:3A}, the comparison between reference and denoised HRTFs in Sec.~\ref{sec:3B}, and the correlation between the geometric metrics evaluated on the point clouds and the metrics computed on the numerical HRTFs related to each ear shape in Sec.~\ref{sec:3C}. The results are further discussed in Sec.~\ref{sec:4}, and the conclusion is provided in Sec.~\ref{sec:5}.

\section{Methods}
\label{sec:2}
%

\subsection{Geometric metrics}
\label{sec:2A}
Various metrics can be used to evaluate the distance between a noisy scan and a reference geometry; e.g.\ their average (\textbf{Avg}) or maximum (\textbf{Max}) absolute Euclidean distance \cite{Dinakaran2018PerceptuallySystems}, or the accuracy (\textbf{Acc}), defined as the $95^{\mathrm{th}}$ percentile of the absolute distance \cite{Reichinger2013EvaluationPinnas}. Completeness (\textbf{Cmp}) and Chamfer Distance (\textbf{CD}), evaluated between reference and photogrammetric pinna point clouds, are shown to be highly negatively and positively correlated with several HRTF metrics, respectively \cite{DiGiusto2023AnalysisHead}. \textbf{Cmp} is defined as the percentage of points ($x$) in the reference point cloud ($X$) at a distance smaller than \SI{1}{\milli\meter} from the points ($y$) of a scanned point cloud ($Y$) \cite{Reichinger2013EvaluationPinnas}, while the \textbf{CD} is defined as:
\begin{align}
\label{eq:CD}
\mathrm{\textbf{CD}} = \frac{\sum\limits_{x \in X} \min\limits_{y \in Y} d(x,y)^2}{N_X} + \frac{\sum\limits_{y \in Y} \min\limits_{x \in X} d(y,x)^2}{N_Y},
\end{align}
with $d(x,y)$ representing the Euclidean distance between the points of $X$ and $Y$, with cardinality $N_X$ and $N_Y$, respectively. Two additional metrics are included in this analysis to further characterise the error in the scans, and assess their relation to the HRTF metrics computed on the acquired geometries. The first is the Hausdorff Distance (\textbf{HD}) \cite{Cignoni1998Metro:Surfaces}, defined as:
\begin{align}
\label{eq:HD}
\mathrm{\textbf{HD}} = \max \left\{\sup_{x \in X} \inf_{y \in Y} d(x,y),\ \sup_{y \in Y} \inf_{x \in X} d(y,x)\right\},
\end{align}
where $\sup$ and $\inf$ denote the supremum and infimum operators, respectively. The second metric takes into account that all photogrammetric scans are related to a known surface mesh. Thus, a distance between point cloud and reference mesh is leveraged, as it avoids the component of the error tangent to the surface, which might affect the metrics computed between two point clouds \cite{Rakotosaona2019PointCleanNet:Clouds}. This is named Mesh Distance (\textbf{MD}) and is defined as:
\begin{align}
\label{eq:MD}
\mathrm{\textbf{MD}} = \frac{\sum\limits_{y \in Y} d_M(y,M)^2}{N_Y},
\end{align}
with $d_M(y,M)$ indicating the distance between the points in $Y$ and the closest face of the mesh $M$ \cite{Cignoni2008MeshLab:Tool}. For an easier assessment of the denoising effects across different tested algorithms, another metric is introduced, namely the Noise Reduction (\textbf{NR}), defined as:
\begin{align}
\label{eq:NR}
\mathrm{\textbf{NR}} = \frac{D(Y,X) - D(\widetilde{Y},X)}{D(Y,X)},
\end{align}
where $D(Y,X)$ and $D(\widetilde{Y},X)$ represent a distance metric computed between the reference geometry, and the original ($Y$) and denoised version ($\widetilde{Y}$) of the scanned point cloud, respectively. The \textbf{NR} is a relative metric, with a positive value indicating effective denoising, e.g.\ $\mathrm{\textbf{NR}} = \SI{100}{\percent}$ corresponds to the complete denoising of the scanned point cloud, while a negative value reflects an increase in error introduced by the denoising algorithm.

\subsection{Ear dataset}
\label{sec:2B}
DNN models require large amounts of data for effective supervised training. Therefore, several scans are carried out on different dummy head geometries. Specifically, three dummy heads are scanned, allowing the use of accurate 3D scanning techniques, such as laser or structured-light scanning, to obtain a reliable approximation of the underlying shape for evaluating the photogrammetric scanning error. The dummy heads used in the current analysis are the Neumann KU100 (\textit{KU1}), the Neutrik CORTEX MK2 (\textit{COR}) and the GRAS KEMAR 45BB (\textit{KEM}), with the latter two also containing a partial and full torso simulator, respectively. A precise scan of the \textit{KU1} is obtained through a line laser scanner on the dismounted parts of the dummy head, merging and meshing the results in a uniform triangular mesh with an Average Element Length (AEL) of \SI{0.6}{\milli\meter} \cite{DiGiusto2023AnalysisHead}. For the \textit{COR}, the full geometry is acquired through a structured-light scanner, in relation to perceptual validation studies of modelled binaural room impulse responses \cite{Rychtarikova2009BinauralRooms, Rychtarikova2011PerceptualUnderstanding}. The resulting mesh shows an AEL of \SI{3.5}{\milli\meter} at the torso and head, and \SI{1.5}{\milli\meter} at the ears, reaching sub-\SI{}{\milli\meter} resolution at the most curved pinna locations. The \textit{KEM} mesh, including the full torso morphology, is obtained from the CAD model of the entire dummy head, by meshing it with uniform triangular elements with an AEL of \SI{0.6}{\milli\meter}. This model contains the right and left anthropometric pinnae, i.e.\ GRAS KB5000 and KB5001, respectively. Furthermore, another set of custom 3D-printed pinnae in two different materials is mounted on the \textit{KEM} \cite{Sinev2023IndividualPinnae}, derived from an individual scan of a subject's morphology up to the eardrum \cite{Roden2020TheResearch}. The reference geometries are obtained by substituting the original \textit{KEM} ears with the custom pinna shapes. All these meshes are aligned as in \cite{DiGiusto2023AnalysisHead}, having the interaural centre at the origin, the $y$-axis passing through the ear canal centres, and the $x$-axis parallel to the Frankfurt plane.

Photogrammetric scans of the dummy heads are carried out using an optical treatment of their surface to obtain the best possible outcome from the photogrammetric reconstruction algorithm. The scans are conducted as in \cite{DiGiusto2023AnalysisHead}, by acquiring videos of the dummy heads with a smartphone camera. The head and each ear are scanned separately for a duration of around \SI{2}{\minute} each, recording them from several angles. Around \SI{100}{} frames are extracted at regular intervals from these videos and used as an input for a photogrammetric reconstruction algorithm \cite{Schonberger2016Structure-from-MotionRevisited, Schonberger2016PixelwiseStereo}. Two scans are carried out on the \textit{KU1}. In the first white matt scanning spray is utilised as the optical treatment, while in the second chalk marks of different colours are randomly drawn on the head and pinnae. A single scan is carried out on the sprayed \textit{COR} dummy head, while for the \textit{KEM} three scans are taken, derived from the sprayed dummy head geometry with original anthropometric ears, and custom 3D-printed ears in white and black resin \cite{Sinev2023IndividualPinnae}.

\begin{figure}[ht]
\centering
\begin{subfigure}[t]{\columnwidth}
\centering
\includegraphics[width=\columnwidth]{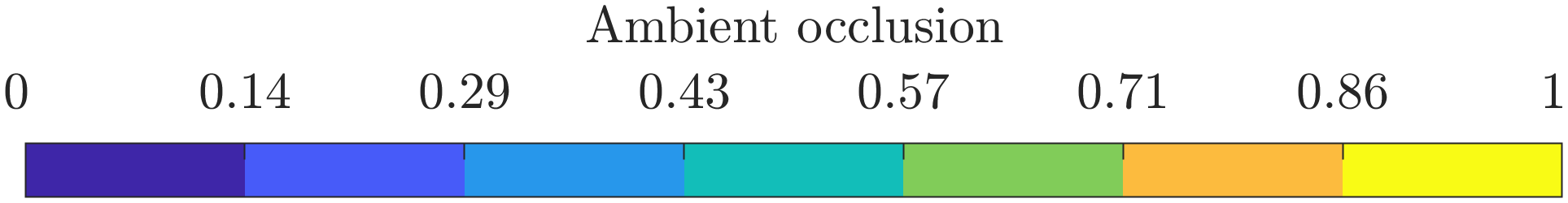}
\end{subfigure} \\
\vskip6pt
\begin{subfigure}[t]{0.48\columnwidth}
\centering
\includegraphics[width=\columnwidth]{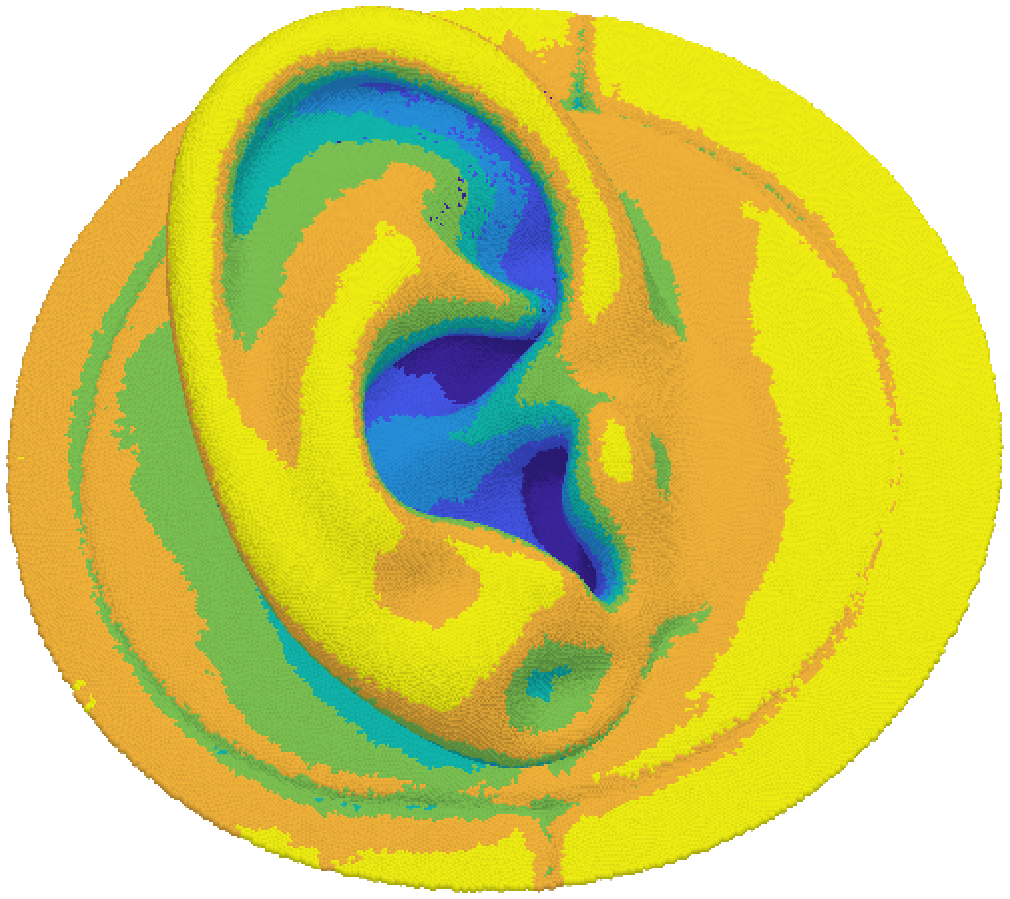}
\end{subfigure} \\
\caption{\textit{KU1} right ear point cloud derived from the laser scan. The colour shows the ambient occlusion at each point.}
\label{fig:Las_Ao}
\end{figure}

The number of available scans and matching reference data for training the DL models is limited. Therefore, the photogrammetric error in the dummy head scans is analysed to extract its inherent characteristics, and artificially corrupt other ear shapes with a synthetic error defined to have similar properties. The raw photogrammetric scans are processed as in \cite{DiGiusto2023AnalysisHead}. The iterative closest point algorithm \cite{Cignoni2008MeshLab:Tool} is leveraged to optimise the scaling and alignment between the scanned pinnae and the reference meshes. The Signed Distance Function (SDF) is evaluated between photogrammetric scans and reference meshes. The SDF presents a positive or negative sign if a point lies outside or inside the surface mesh, respectively \cite{Cignoni1998Metro:Surfaces}, allowing to assess the direction of the photogrammetric error. The SDF distribution is evaluated on each scanned ear, and the best fitting distribution to the observed one is found \cite{Taskesen_distfit_is_a_2020}, i.e.\ the $t$-distribution with median parameters $\nu_\mathrm{med} = \SI{1.95}{}$, $\mu_\mathrm{med} = \SI{0.02}{\percent}$ of $l$ and $\sigma_\mathrm{med} = \SI{0.20}{\percent}$ of $l$, where $\nu$, $\mu$ and $\sigma$ indicate the shape, location and scale parameters of the $t$-distribution, respectively, while $l$ represents the point cloud bounding box diagonal length, having a median value around \SI{117}{\milli\meter}. Additionally, the locations with the highest photogrammetric error are approximated using the Ambient Occlusion (AO), based on the observation that the scanning error is higher for decreasing visibility, which could be estimated through this metric \cite{Reichinger2013EvaluationPinnas}. The AO is calculated on each point of the pinna geometry by determining the fraction of directions over a hemisphere centred at that point which are unobstructed by the surface itself, and is expressed as a value between \SI{0}{} and \SI{1}{}. This metric, computed on the laser-scanned \textit{KU1} right ear geometry, is shown in Fig.~\ref{fig:Las_Ao}.

The complement of AO, i.e.\ $\mathrm{AO}^c = 1 - \mathrm{AO}$, is used to have high values at the most occluded points. To further target the most concave pinna parts, where the photogrammetric error tends to be focused, $\mathrm{AO}^c$ is raised to a power determined by finding the highest Pearson's correlation coefficient between the absolute SDF amplitude at each ear position and the $\mathrm{AO}^c$ raised to an arbitrary positive integer power. The results indicate that the best correlation, evaluated as the median over all scans, is achieved with $(\mathrm{AO}^c)^3$. The \textbf{Cmp} is also analysed, as it relates to the presence of holes in the photogrammetric geometries. Its median reaches a value of $\mathrm{\textbf{Cmp}}_\mathrm{med} = \SI{93.45}{\percent}$ across the acquired ears.

Several individual meshes, derived from a public database containing head and ear geometries of real subjects acquired with a structured-light scanner, namely the HUTUBS HRTF database\footnote{The HUTUBS HRTF database. Accessed: November 20, 2022. \url{https://doi.org/10.14279/depositonce-8487}} \cite{Brinkmann2019AResponses}, are corrupted using a synthetic error with a distribution similar to that observed in the dummy head photogrammetric scans. This process follows a specific procedure. First, the meshes in the database are uniformly remeshed to an average edge length of \SI{0.3}{\milli\meter}, and the AO is computed on them \cite{Jacobson2018libigl:Library}. Next, left ear point clouds are extracted by cutting out sections of the mesh vertices with a circular geometry of \SI{40}{\milli\meter} radius, perpendicular to the $y$-axis and centred at the ear canal, also storing the AO at each of the extracted points. This yields pinna geometries containing around \SI{100000}{} points, in line with the photogrammetric ear scans. A subsample of these points is removed based on the \textbf{Cmp} values observed in the dummy head scans, achieved by randomly sampling the extracted ear point clouds with a distribution proportional to $1 - (\mathrm{AO}^c)^3$. Consequently, points at visible locations are more likely to be sampled than highly occluded ones; thereby mimicking the holes primarily found in the photogrammetric pinna cavities. Subsequently, a random realisation of the $t$-distribution ($t$) is generated, with matching number of samples to the subsampled ear shapes. Given the high cardinality of the point clouds, some samples of $t$ ($t_i$) may exhibit extreme values. Since this is not observed in the real scans, the median value of the maximum absolute SDF evaluated on them ($\mathrm{\textbf{Max}}_\mathrm{med}$) is used to obtain a scaled distribution ($\hat{t}$), whose samples ($\hat{t}_i$) are defined as:
\begin{align}
\label{eq:t}
\hat{t}_i = \begin{cases} 
t_i & \text{if } \lvert t_i \rvert \leq \mathrm{\textbf{Max}}_\mathrm{med},\\
t_i \cdot \frac{\mathrm{\textbf{Max}}_\mathrm{med}}{\max\lvert t \rvert} & \text{if } \lvert t_i \rvert > \mathrm{\textbf{Max}}_\mathrm{med},
\end{cases}
\end{align}
with $\mathrm{\textbf{Max}}_\mathrm{med}=\SI{3.78}{\percent}$ of $l$. The $\hat{t}$ values obtained through Eq.~(\ref{eq:t}) are then applied to displace the ear point cloud samples ($x_i$), generating points affected by synthetic error ($y_i$) according to:
\begin{align}
\label{eq:y_i}
y_i = x_i + \hat{t}_i \cdot n_i,
\end{align}
where $n_i$ indicates the point normal related to $x_i$. To achieve a result resembling a photogrammetric scan, the ear points are first arranged in ascending order based on their $(\mathrm{AO}^c)^3$, then the $\hat{t}_i$ related to each $x_i$ in Eq.~(\ref{eq:y_i}) is iteratively selected from $\hat{t}$ using a random choice algorithm with a distribution proportional to the flipped $\lvert \hat{t} \rvert$. Consequently, points at the most occluded ear locations, i.e.\ those with the highest $(\mathrm{AO}^c)^3$, are more likely to be displaced by the highest absolute values in $\hat{t}$, and vice versa. Various levels of noise are created by modifying the $\sigma$ parameter within the range from $\sigma_{\min} = \SI{0.1}{\percent}$ to $\sigma_{\max} = \SI{0.5}{\percent}$ of $l$, while the \textbf{Cmp} value is chosen depending on $\sigma$, according to:
\begin{align}
\label{eq:cmp_rng}
\mathrm{\textbf{Cmp}} &= \mathrm{\textbf{Cmp}}_{\max} \cdot (1 - \eta) + \mathrm{\textbf{Cmp}}_{\min} \cdot \eta,\\
\label{eq:eta}
\eta &= \frac{\sigma - \sigma_{\min}}{\sigma_{\max} - \sigma_{\min}},
\end{align}
with $\mathrm{\textbf{Cmp}}_{\max} = \SI{98}{\percent}$ and $\mathrm{\textbf{Cmp}}_{\min} = \SI{80}{\percent}$. Therefore, if $\sigma = \sigma_{\min}$ in Eq.~(\ref{eq:eta}), then $\mathrm{\textbf{Cmp}} = \mathrm{\textbf{Cmp}}_{\max}$ in Eq.~(\ref{eq:cmp_rng}), representing a scan with low photogrammetric error, and vice versa. The ranges of $\sigma$ and \textbf{Cmp} are chosen to approximate the extreme values of these parameters observed in the acquired dummy head photogrammetric scans.

\begin{figure}[ht]
\centering
\begin{subfigure}[t]{\columnwidth}
\centering
\includegraphics[width=\columnwidth]{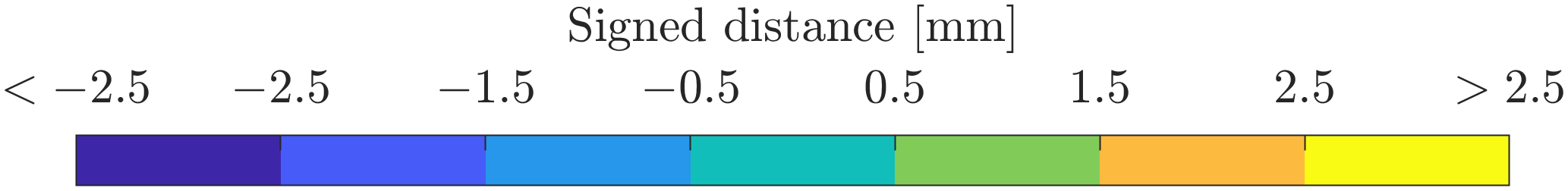}
\end{subfigure} \\ 
\vskip6pt
\begin{subfigure}[t]{0.48\columnwidth}
\centering
\includegraphics[width=\columnwidth]{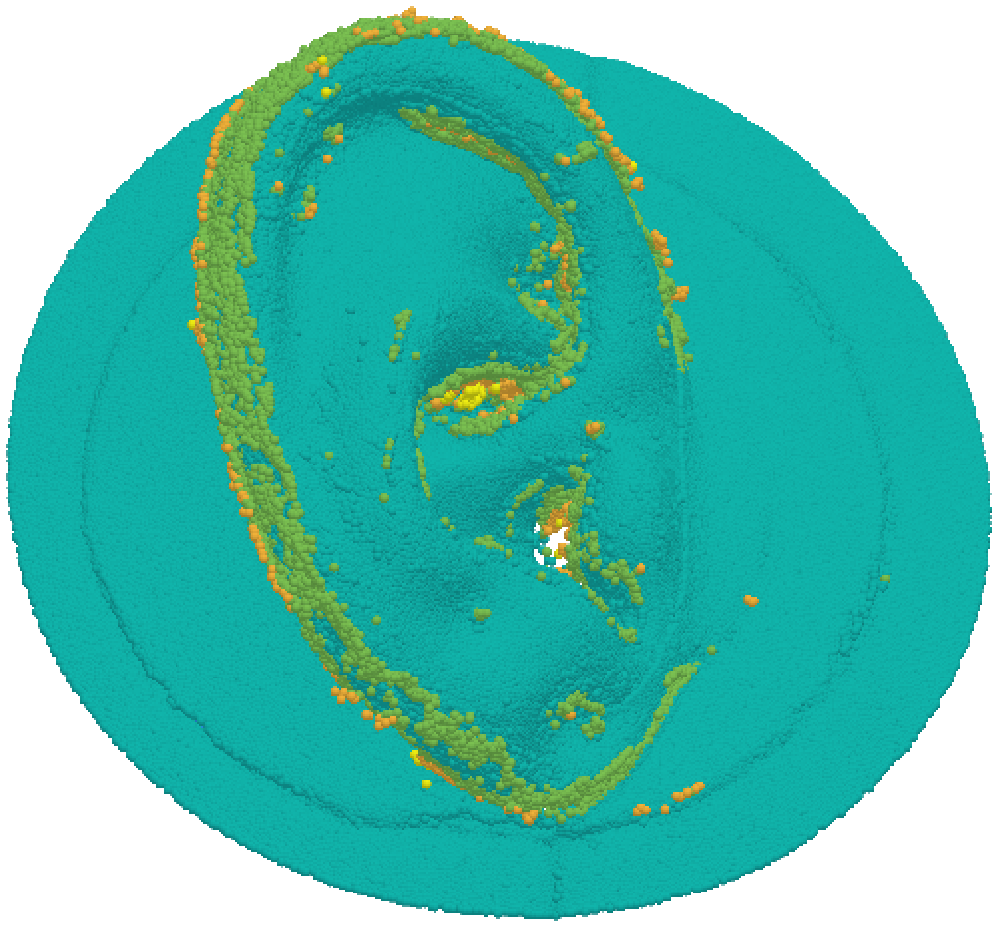}
\caption{Chalk marks.}
\label{fig:Pho_Raw_a}
\end{subfigure}\hfill 
\begin{subfigure}[t]{0.48\columnwidth}
\centering
\includegraphics[width=\columnwidth]{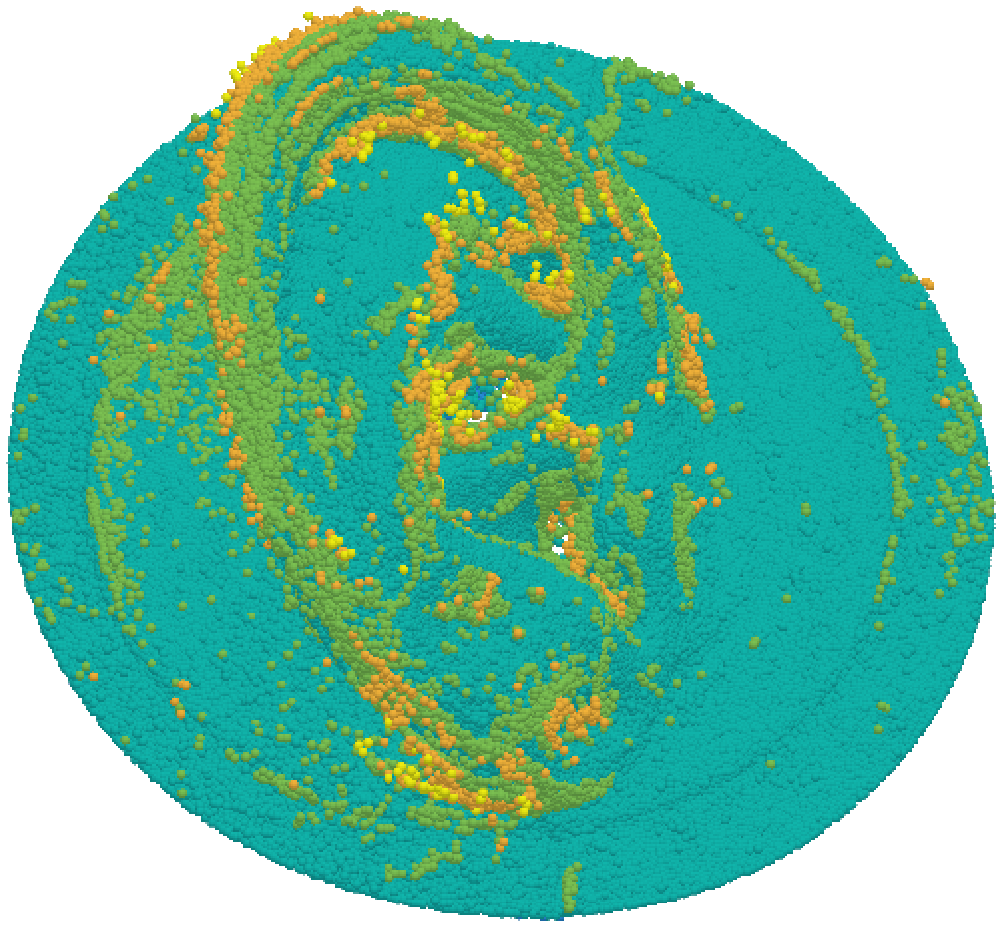}
\caption{Scanning spray.}
\label{fig:Pho_Raw_b}
\end{subfigure} \\
\begin{subfigure}[t]{0.48\columnwidth}
\centering
\includegraphics[width=\columnwidth]{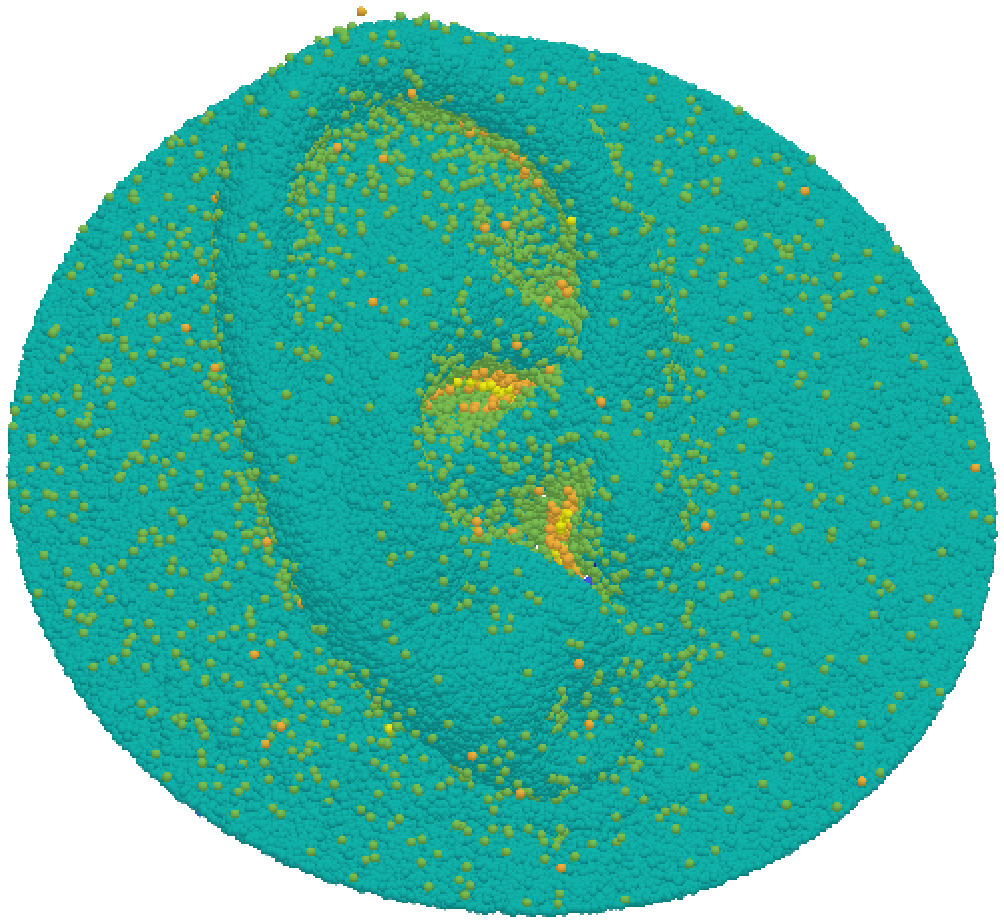}
\caption{Synthetic error.}
\label{fig:Pho_Raw_c}
\end{subfigure}\hfill 
\begin{subfigure}[t]{0.48\columnwidth}
\end{subfigure} \\
\caption{\textit{KU1} right ear point clouds derived from the photogrammetric scan with chalk marks (a) and scanning spray (b), and laser scan corrupted with synthetic photogrammetric error (c). The colour shows the signed distance from the laser scan, cropped at $\pm\SI{2.5}{\milli\meter}$.}
\label{fig:Pho_Raw}
\end{figure}

The left ears of $20$ different subjects are selected as training data; to these shapes, $6$ levels of synthetic photogrammetric error are applied, i.e.\ $\sigma$ values of \SIlist{0.1; 0.2; 0.3; 0.4; 0.5}{\percent} of $l$, also including the original unmodified points, as done in \cite{Rakotosaona2019PointCleanNet:Clouds}, to train the network to preserve the clean geometry. The validation and testing datasets consist of the left ears of $10$ subjects each, corrupted with $3$ levels of synthetic photogrammetric error, i.e.\ $\sigma$ values of \SIlist{0.1; 0.3; 0.5}{\percent}. These choices yield $120$ point clouds for training and $30$ samples each for validation and testing. For each noisy point cloud, a matching clean one is extracted from the original mesh and stored without modifications to create noisy-clean data pairs for supervised training. Additionally, a dataset to evaluate the denoising effect on the experimental scanning error is created by extracting ear point clouds from the dummy head photogrammetric scans without further modifications. The AO for each point is obtained from the closest vertex in the related reference mesh. This dataset includes both left and right ear point clouds, with the latter being mirrored to the left to be in line with the synthetic training data, resulting in $12$ ear shapes from the $6$ dummy head scans. The similarity between real and synthetic error is estimated by computing the Pearson's correlation coefficient of the SDF for each photogrammetric scan and that of a realisation of the modelled error on the same ear geometry. This coefficient reaches a median value of \SI{0.48}{}, while ranging from \SI{0.21}{} to \SI{0.63}{}, indicating a moderate positive correlation. Although relatively low, these values are considered acceptable given the inherently random nature of the real scanning error and the tested synthetic error realisation. The lowest correlation is observed is scans presenting several holes, likely due to the ineffectiveness of the white scanning spray against the light colour of the dummy head material, e.g.\ in the \textit{COR} scan. Figure~\ref{fig:Pho_Raw} visualises the real and synthetic photogrammetric error on the right ear of the \textit{KU1}; specifically, Fig.~\ref{fig:Pho_Raw_a} and \ref{fig:Pho_Raw_b} show the raw outcomes of the scans with chalk marks and scanning spray, respectively, while Fig.~\ref{fig:Pho_Raw_c} displays a realisation of the synthetic error. Although the real and synthetic error exhibit some similarities, they show different trends, especially at the most convex pinna parts. Nonetheless, the concentration of geometric deviation at the concave pinna locations, considered to be the most detrimental characteristic of the photogrammetric error in relation to HRTFs computed on the scanned geometries \cite{DiGiusto2023AnalysisHead}, is effectively captured in the modelled ear scan.

A subset of the scanned dummy head geometries is further used to train a modified version of a DNN. This aims to assess whether fine-tuning on a limited set of scanned ear shapes can enhance the model's denoising performance, given the only moderate correlation observed between real and synthetic scanning error. The $12$ scanned dummy head ear point clouds are divided into a training set, containing the $3$ left and right ear scans of the \textit{KEM}, and a validation set comprising the \textit{COR} ears, while the \textit{KU1} point clouds are reserved for testing. Additionally, to increase the number of geometries on which the denoising is tested and the HRTFs are computed, the $2$ scans conducted on the \textit{KU1} are repeated using the same optical treatments. Particular attention is given to scanning the concave pinna structures in the repeated scans, especially the cymba conchae and fossa triangularis, given their strong influence on the HRTF spectral features \cite{DiGiusto2023AnalysisHead}. This is done by placing greater emphasis on the optical treatment application, as it is more challenging at the most concave-shaped areas, and focusing the second minute of the video recording on these locations, capturing them from different angles and at a distance of approximately \SI{0.2}{\meter}. It is important to note that these latter scans were taken by the same operator after the creation of the synthetic ear dataset; therefore, their statistics are not considered in the dataset creation. While also including these results might alter the median values used for synthetic error generation, their contribution is expected to be minimal due to the similarity with the previous scans. Indeed, when including the two additional scans, the median values are $\nu_\mathrm{med} = \SI{1.81}{}$, $\mu_\mathrm{med} = \SI{0.02}{\percent}$ of $l$ and $\sigma_\mathrm{med} = \SI{0.16}{\percent}$ of $l$, reflecting only minor deviations from the previous results.

\subsection{Denoising models}
\label{sec:2C}
Three DL architectures for point cloud denoising are tested, i.e.\ PointCleanNet (\textit{PCN}) \cite{Rakotosaona2019PointCleanNet:Clouds}, DMRDenoise (\textit{DMR}) \cite{Luo2020DifferentiableDenoising}, and Score-Denoise (\textit{SCR}) \cite{Luo2021Score-BasedDenoising}. These models operate on local point cloud patches centred at query points, making them efficient and robust for estimating local properties of the underlying surfaces. The hyperparameters used for training each algorithm adhere to the original specifications, unless otherwise specified.

\textit{PCN} is based on a PCPNet architecture \cite{Guerrero2018PCPNet:Clouds} and incorporates spatial transformer networks, feature extractors, symmetric pooling operators, and regressor networks. It leverages features from patches containing points within a given maximal distance from the query point to estimate an optimal displacement vector for the latter. The model is trained in a supervised manner on noisy-clean data pairs, where the noisy point clouds are artificially corrupted with varying levels of Gaussian noise. The loss function comprises the squared $L_2$ norm between noisy and clean points, along with a regularisation term to promote uniform point distribution. Training across multiple noise levels enables the model to handle different noise magnitudes, allowing an iterative application where the denoised point cloud is re-fed to the model to further reduce residual noise \cite{Rakotosaona2019PointCleanNet:Clouds}.

\textit{DMR} adopts an autoencoder-like architecture, including a representation encoder and a manifold reconstruction decoder. The encoder leverages a differentiable down-sampling unit trained to sample a subset of the point cloud closer to the underlying geometry, employing several feature extraction units composed of multiple dynamic graph convolution layers, assembled with different k-nearest-neighbour values to concatenate multi-scale features. These features are fed to a differentiable pooling operator selecting points based on their proximity to the underlying manifold. The decoder, implemented by a Multi-Layer Perceptron (MLP), is used to transforms each sampled point and its neighbours in a patch manifold, on which the denoised point cloud is up-sampled. The model is trained in a supervised manner using two loss functions, i.e.\ the \textbf{CD} to assess the quality of the down-sampling and the Earth Mover's Distance (\textbf{EMD}) to quantify the distance between the denoised point cloud and the ground truth points \cite{Luo2020DifferentiableDenoising}. Given that the \textbf{EMD} requires point clouds of equal size, to fine-tune this model on the ear dataset the \textbf{CD} is used also as the second loss function. This is done since the point cloud subsampling, mimicking the incompleteness affecting the photogrammetric scans, might relate to noisy-clean data pairs containing different numbers of points.

\textit{SCR} is designed to learn the score of the underlying error distribution in the noisy point cloud, assumed to have a continuous Probability Density Function (PDF) with a unique mode at \SI{0}{}. Under these assumptions, the noisy points' PDF can be expressed as the convolution of the clean points' PDF and the error PDF, and denoising can be achieved through gradient ascent on the score of the noisy points PDF. The network is composed of a feature extraction unit, entailing a stack of densely connected dynamic graph convolution layers which extracts local and non-local features for each point, and a score estimator, modelled as an MLP, which estimates the score function at each point. The supervised training aligns the estimated score with the ground truth score, using an objective function matching the score at the noisy point position and in its neighbouring area, given that the denoising displaces the points during gradient ascent \cite{Luo2021Score-BasedDenoising}.

The DNN models are compared against a non-DL based denoising approach, specifically a polynomial filtering method (\textit{POL}). This technique involves an initial implicit plane fitting on the 3D position of a local query point neighbourhood, and a subsequent approximation of the neighbourhood by a 2D polynomial of arbitrary order. This provides a reasonable estimation of the surface normal, which is then used to project the query point onto the fitted polynomial surface \cite{Egner2023PolynomialPlates}. In this study, the local point neighbourhood is defined as a sphere with \SI{3}{\milli\meter} radius, and a 2\textsuperscript{nd} order polynomial approximation is employed.

\begin{figure}[ht]
\centering
\includegraphics[width=\columnwidth]{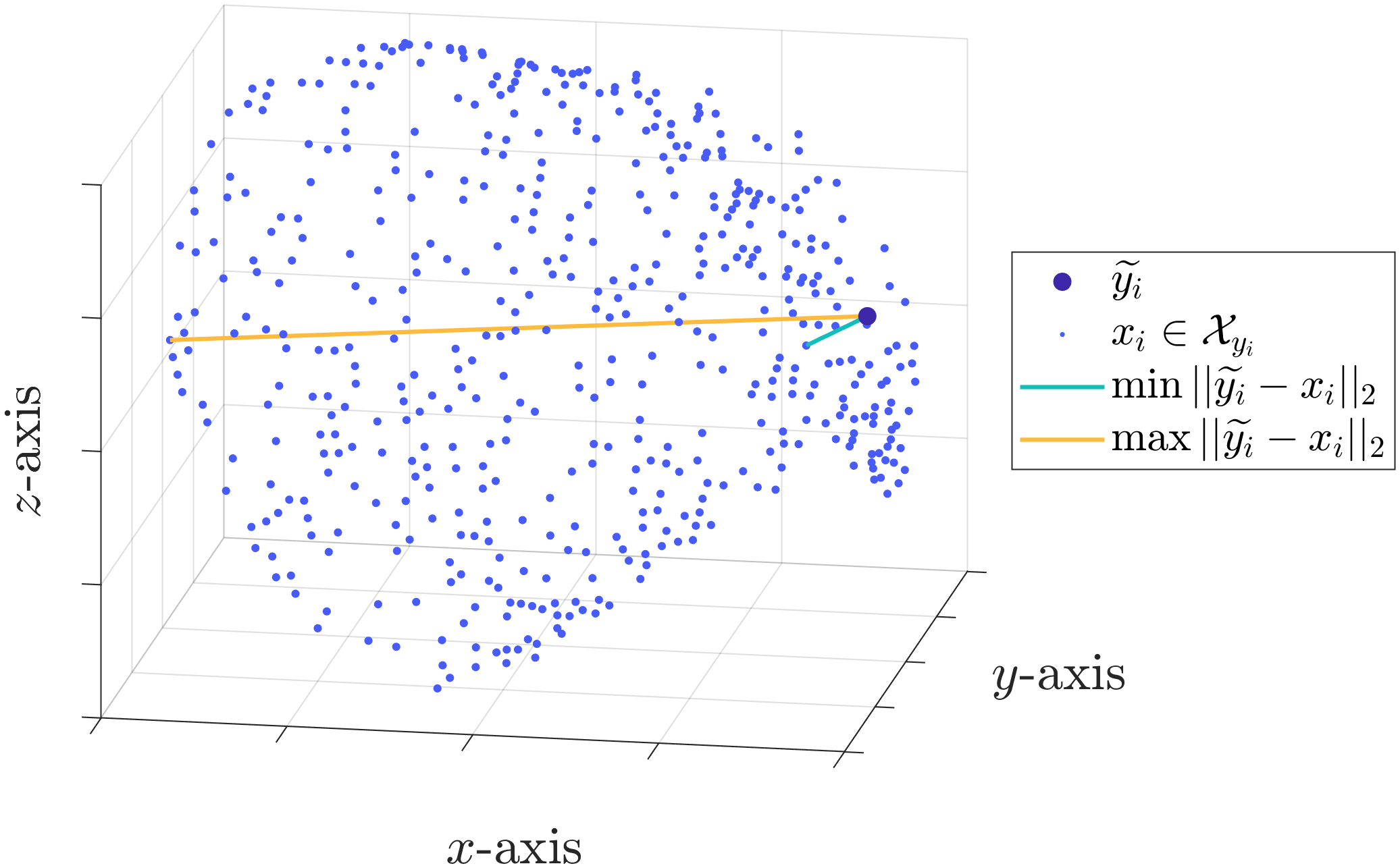}
\caption{Example of denoised query point ($\widetilde{y_i}$) and points ($x_i$) of its related local reference patch ($\mathcal{X}_{y_i}$). The lines show the minimum and maximum $L_2$ norm between $\widetilde{y_i}$ and $\mathcal{X}_{y_i}$, used in the loss function ($L$).}
\label{fig:loss}
\end{figure}

\textit{PCN} is further modified to improve its effectiveness in denoising ear point clouds affected by photogrammetric error; this modified version of \textit{PCN} is referred to as \textit{PCNm}. 
Similarly to the original \textit{PCN} loss function, the \textit{PCNm} loss ($L$) is also defined as a weighted combination of two terms:
\begin{align}
L = \alpha \cdot \widehat{L_s} + (1 - \alpha) \cdot L_r,
\label{eq:l_t}
\end{align}
with $\alpha$ being a weighting factor, $\widehat{L_s}$ representing the distance between each denoised point and the clean point cloud, and $L_r$ a regularisation term to promote a uniform distribution of the denoised point cloud across the entire surface. $\widehat{L_s}$ is expressed as:
\begin{align}
\widehat{L_s} = \min\limits_{{x_i} \in \mathcal{X}_{y_i}} \lVert \widetilde{y_i} - x_i \rVert_2^2 \cdot \widehat{\mathrm{AO}^c_{y_i}},
\label{eq:l_s}
\end{align}
where $\lVert \widetilde{y_i} - x_i \rVert_2$ is the $L_2$ norm between the denoised query point $\widetilde{y_i}$, and the points $x_i$ of the local reference patch $\mathcal{X}_{y_i}$ centred at $y_i$. The term $\widehat{\mathrm{AO}^c_{y_i}}$ represents the $\mathrm{AO}^c$ at $y_i$ normalised by its average value across the full point cloud. The $\alpha$ and $L_r$ terms in Eq.~(\ref{eq:l_t}) follow the original \textit{PCN} formulation, i.e.\ $\alpha = \SI{0.99}{}$ and $L_r = \max_{{x_i} \in \mathcal{X}_{y_i}} \lVert \widetilde{y_i} - x_i \rVert_2^2$ \cite{Rakotosaona2019PointCleanNet:Clouds}. The factor proportional to $\mathrm{AO}^c$ in Eq.~(\ref{eq:l_s}) focuses the denoising on the pinna cavities, as the HRTF spectral features are highly sensitive to geometric error in these ear regions \cite{DiGiusto2023AnalysisHead, Stitt2021SensitivityModel, Ghorbal2017PinnaSets}. A visualisation of the denoised query point and related local reference patch, i.e.\ points within $\SI{5}{\percent}$ of $l$ from the query, alongside their minimum and maximum distances, is presented in Fig.~\ref{fig:loss}. 

The architecture of \textit{PCNm} is also modified by including a global subsample of the ear in the DNN input, in addition to the local point patch, as in \cite{Erler2020Points2SurfClouds}. This aims to aid the spatial transformer networks in learning the canonical orientation of the input points prior to denoising, while also providing information on the overall shape. The global subsample points are selected from the full input point cloud using a weight directly proportional to their distance from the query point, ensuring that points closer to the latter are more likely to be included. It should be noted that the global point patch does not affect the loss function, which is computed solely on points of the local input patch.

\subsection{HRTF computation and assessment}
\label{sec:2D}
The process and parameters used to generate suitable meshes for HRTF computation from reference and photogrammetric scans adhere to those outlined in \cite{DiGiusto2023AnalysisHead}. A similar approach is applied to the denoised point clouds, with the only difference being that the ears from the original photogrammetric head scan are removed and replaced by the denoised geometries. It should be noted that the right ear point clouds are first mirrored to the left to align with the training data, denoised, and then mirrored back to the right. The results are processed to obtain watertight triangle meshes with an AEL of \SI{0.6}{\milli\meter}. Additionally, the ear canals of each mesh are cut on a plane perpendicular to the $y$-axis, at a distance of $\pm\SI{68.5}{\milli\meter}$ from the interaural centre, and closed with a planar surface, ensuring the ear canal centre is consistently defined at a common location. Pymeshlab \cite{Muntoni2021PyMeshLab} and OpenFlipper \cite{Mobius2010OpenFlipper:Framework} are used to perform these processing steps. To accelerate the HRTF computation, a grading algorithm is applied to the uniform meshes \cite{Ziegelwanger2016AFunctions}; the final graded head meshes feature element sizes ranging from \SI{1}{\milli\meter} near one ear canal's centre to \SI{15}{\milli\meter} on the opposite side. 

A software utilising FEM Adaptive Order is chosen for the HRTF computation \cite{Beriot2016EfficientProblems}, where the order of the finite elements is selected according to an a priori error estimator tuned to fine accuracy level. Non-reflective boundary conditions are implemented through an Automatically Matched Layer (AML) \cite{Beriot2020AnShape}, defined on a convex mesh surrounding the head geometry, at a distance of \SI{30}{\milli\meter} and with an AEL of \SI{15}{\milli\meter}. Sound hard boundary conditions are applied on the full dummy head geometry, given the unknown impedance of its material, and the low perceptual effect observed when including measured boundary impedance values in simulations conducted on 3D-printed ear replicas \cite{DiGiusto2023AnalysisReplicas}. The outcome is evaluated on a spherical grid having a radius of \SI{1.2}{\meter}, and an angular resolution of \SI{2.5}{\degree} in azimuth ($\theta$) and \SI{5}{\degree} in elevation ($\varphi$). All other simulation parameters are kept the same as in \cite{DiGiusto2023AnalysisHead}. Additionally, the post-processing steps used to convert simulation results into Head-Related Impulse Responses (HRIRs) are identical to those outlined in \cite{DiGiusto2023AnalysisHead}. These also include the estimation of the frequency scaling factor ($\alpha$) to compensate for potential deviations between numerical and experimental data \cite{Jin2014CreatingDatabase}, and the conversion in Common Transfer Functions and Directional Transfer Functions (DTFs) \cite{Brinkmann2017AOrientations}.

\begin{figure}[ht]
\centering
\includegraphics[width=\columnwidth]{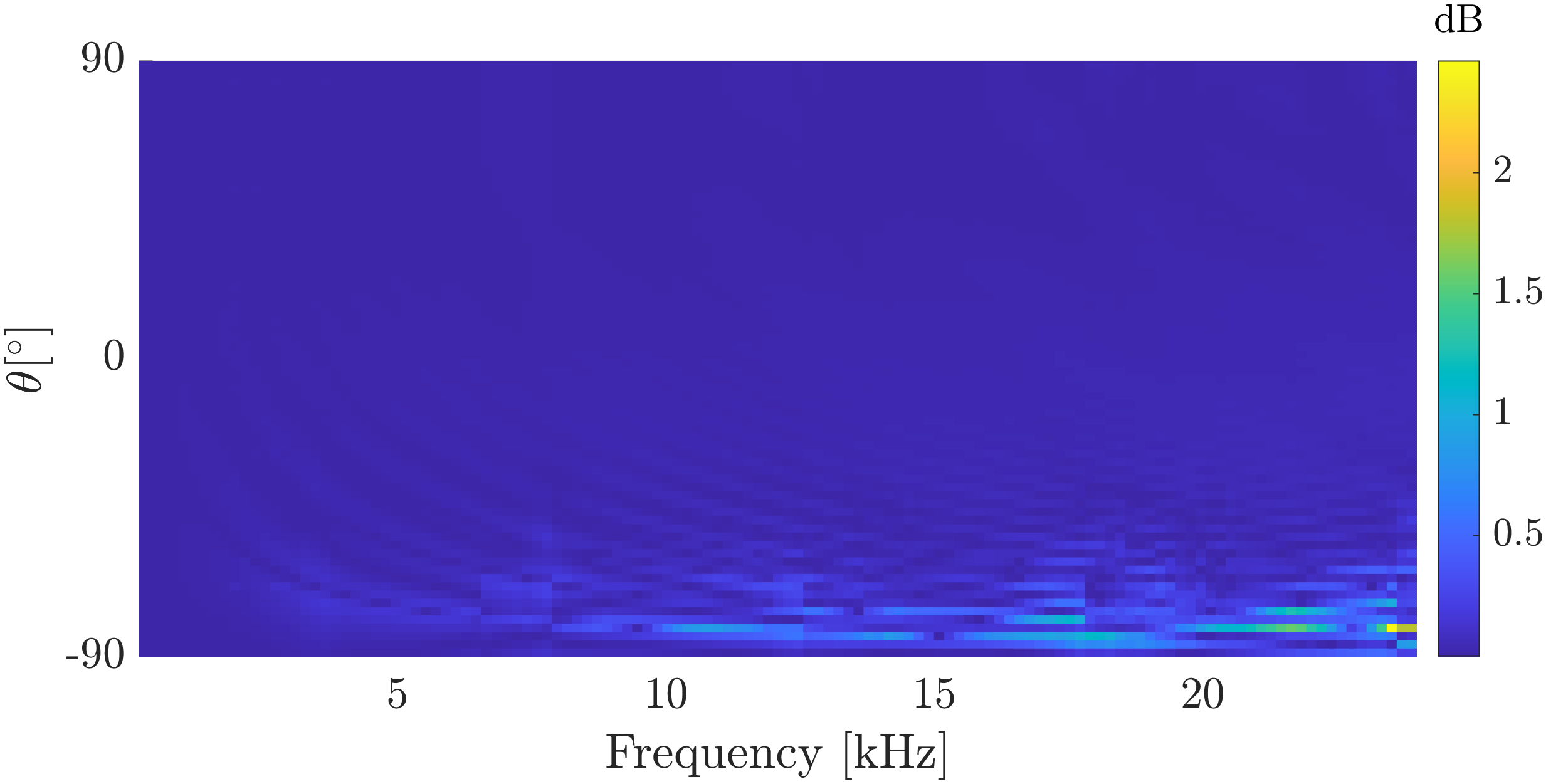}
\caption{Absolute magnitude difference between analytical and FEM Adaptive Order solutions for the scattering of a point source by a rigid sphere. The results are shown on the horizontal plane from contralateral to ipsilateral side, centred at $\theta = \SIlist{-90; 90}{\degree}$, respectively.}
\label{fig:sph_ref}
\end{figure}

To validate the outcome of the employed FEM software, a test is carried out on a simplified case for which a benchmark analytical solution is available, i.e.\ the scattering of a point source by a rigid sphere \cite{Jacobsen2013FundamentalsAcoustics}. The simulation parameters are kept the same as those used for the HRTF computations, but the head geometry is substituted with a graded spherical mesh with a radius of \SI{0.1}{\meter}. Given the spherical symmetry of this simplified case, the absolute magnitude difference in \SI{}{\deci\bel} between the analytical and FEM results is plotted for half of the horizontal plane in Fig.~\ref{fig:sph_ref}. Since the source is defined on the left side of the sphere, i.e.\ at $\theta = \SI{90}{\degree}$, the displayed results range from contralateral to ipsilateral  side, centred at $\theta = \SIlist{-90; 90}{\degree}$, respectively.

The outcome shows low errors, reaching a maximum level of \SI{0.05}{\deci\bel} on the ipsilateral side and \SI{2.5}{\deci\bel} on the contralateral side. This is lower than the error reported for the same case in a computation carried out with state-of-the-art BEM software, i.e.\ below \SIlist{1; 6}{\deci\bel} on the ipsilateral and contralateral side, respectively \cite{Pollack2023SpectralEardrum}. As in that study, this error is considered to be negligible, given the low contribution of the contralateral ear to sound localisation performance.

The divergence in binaural metrics between the HRTFs computed on reference and photogrammetric scans is not analysed, as only small deviations, close to the just-noticeable difference in anechoic conditions, are reported between the ITDs and ILDs derived from laser and photogrammetric \textit{KU1} scans \cite{DiGiusto2023AnalysisHead}. The current analysis focuses on elevation perception, for which monaural spectral cues are crucial. The Inter-Subject Spectral Difference (ISSD) is used as an objective metric for comparing the spectra of two different HRTFs \cite{Denk2018SpectralEars}. This is defined as:
\begin{align}
\mathrm{ISSD}_{i,j} = \frac{\sum_{\Psi} \mathrm{Var}_{f} \left(\hat{H}_i(\theta, \phi, f_k) - \hat{H}_j(\theta, \phi, f_k) \right)}{N_\Psi}, \quad
\label{eq:ISSD}
\end{align}
with $\mathrm{Var}_{f}$ indicating variance averaged over frequency ($f$) of $\hat{H}$, representing the \SI{}{\deci\bel} amplitude of gammatone filtered DTFs with 1 Equivalent Rectangular Bandwidth related to each centre frequency $f_k$. The results are averaged across both ears and the number $N_\Psi$ of incidence angles $\Psi = (\theta, \varphi)$.

The ISSD tends to show a strong correlation with perceptually inspired metrics \cite{DiGiusto2023AnalysisHead, Stitt2021SensitivityModel}, generally obtained through a sagittal plane localisation model \cite{Baumgartner2014ModelingListeners} implemented in the Auditory Modeling Toolbox (AMT) version 1.5 \cite{Majdak2022AMTModeling}. In this model, one template and one target DTF are used to estimate the localisation error of a virtual subject, represented by the template DTF, localising sound filtered with another set of DTFs, i.e.\ the target DTF. The results are expressed in terms of Quadrant Error (QE) and Polar Error (PE), representing the percentage of responses falling in a different quadrant from the true source position and the local angular accuracy within the same quadrant, respectively. The sensitivity parameter ($S$), modelling the high inter-individual differences observed in real localisation experiments, is set to $S = \SI{0.21}{}$, modelling the best localiser in an experimental setting \cite{Baumgartner2014ModelingListeners}. Analyses conducted in \cite{DiGiusto2023AnalysisHead} identify several ranges of QE and PE to represent expected localisation errors among groups of individual accurate, inaccurate, and non-individual measured DTFs, derived from HRTF databases of dummy heads, i.e.\ the Club Fritz HRTF database\footnote{The Club Fritz HRTF database. Accessed: April 25, 2022. \url{https://sofacoustics.org/data/database/clubfritz/}}, and human subjects, i.e.\ the ARI HRTF database\footnote{The ARI HRTF database. Accessed: April 25, 2022. \url{https://sofacoustics.org/data/database/ari/}}. The first group includes DTFs accurately acquired on the \textit{KU1}, while the second relates to data affected by measurement error capable of hindering the localisation cues, resulting in higher QE and PE. The last group represents errors obtained when localising sound filtered with non-individual DTFs. These ranges are employed to assess the denoising effect on the DTFs computed from the processed photogrammetric scans. The same parameters as those in \cite{DiGiusto2023AnalysisHead} are used to run the sagittal localisation model analysis.

\section{Results}
\label{sec:3}
%

\subsection{Denoising}
\label{sec:3A}
\begin{table*}[ht]
\centering
\caption{Geometric error metrics of \textit{KU1} ear point clouds derived from the photogrammetric scans. Median values evaluated on the original, repeated, and all scans, considering the full ear or the front part alone. The reference geometry is derived from the \textit{KU1} laser scan.}
\label{tab:scan_metrics}
\vskip3pt
\begin{tabular}{cccccccc}
\hline\hline
ID &
\begin{tabular}{c}\textbf{Acc}\\$[$\SI{}{\milli\meter}$]$\end{tabular} &
\begin{tabular}{c}\textbf{Cmp}\\$[$\SI{}{\percent}$]$\end{tabular} &
\begin{tabular}{c}\textbf{Avg}\\$[$\SI{}{\milli\meter}$]$\end{tabular} &
\begin{tabular}{c}\textbf{Max}\\$[$\SI{}{\milli\meter}$]$\end{tabular} &
\begin{tabular}{c}\textbf{CD}\\$[$\SI{}{\milli\meter\squared}$]$\end{tabular} &
\begin{tabular}{c}\textbf{HD}\\$[$\SI{}{\milli\meter}$]$\end{tabular} &
\begin{tabular}{c}\textbf{MD}\\$[$\SI{}{\milli\meter\squared}$]$\end{tabular}\\
\hline
original full &1.20 &97.0 &0.38 &4.46 &0.49 &4.81 &0.30\\
original front &1.26 &94.8 &0.41 &2.00 &0.53 &3.10 &0.53\\
\hline
repeated full &0.68 &93.6 &0.22 &3.73 &0.41 &4.39 &0.09\\
repeated front &0.67 &94.6 &0.21 &2.00 &0.26 &2.20 &0.26\\
\hline
all full&0.84 &95.7 &0.25 &3.95 &0.48 &4.46 &0.12\\
all front &0.91 &94.6 &0.26 &2.00 & 0.37 &2.93 &0.34\\
\hline\hline
\end{tabular}
\end{table*}
The \textit{KU1} scans are evaluated using the geometric metrics outlined in Sec.~\ref{sec:2A}, with the laser-scanned geometry serving as the reference. These metrics are computed for both the full ear point clouds and the frontal part of the pinna alone, i.e.\ the aspect visible to an external observer looking along the interaural axis, as the latter is considered to be the most acoustically relevant part \cite{Reichinger2013EvaluationPinnas}. Extracting the front of the ear from a point cloud is not trivial due to the unstructured nature of this format and the complex pinna geometry. This is practically achieved by manually extracting the left and right pinna fronts from the laser-scanned \textit{KU1} mesh, and selecting point cloud samples within \SI{2}{\milli\meter} from them. The median value for each metric across all scans is reported in Tab.~\ref{tab:scan_metrics}, also separately including the results from the original and repeated scans.

Comparing the outcomes of the original and repeated scans, it can be seen that most metrics show improved results in the latter, likely stemming from better scanning conditions and increased expertise of the operator. The only metric that worsens in the repeated scans is \textbf{Cmp}; however, for the frontal points, small differences are noted between the \textbf{Cmp} of original and repeated scans. In the original scan, all metrics except \textbf{Max} and \textbf{HD} indicate a greater distance for the front, suggesting larger scanning errors in this region, as contributions from points on the head surface surrounding the ear, which are easier to scan given the flatter shape, are not considered. Nonetheless, the \textbf{HD} tends to show smaller values for the frontal part, implying that the maximum deviation may occur at the back of the ear. This could, however, be an artefact of the procedure used to select the frontal pinna points, i.e.\ points having $\mathrm{\textbf{Max}} \leq \SI{2}{\milli\meter}$ from the front ear mesh, reflecting the common \textbf{Max} value of \SI{2}{\milli\meter} observed across all frontal results. Conversely, when comparing full and frontal point cloud distance metrics from the repeated scans, lower values are seen for the front, except for \textbf{MD}, which follows a trend consistent with the original scan results. This suggests improved scanning outcomes in the frontal part of the pinna for the repeated acquisition. The combined results of original and repeated scans indicate that most metrics exhibit worse values for the frontal part, apart from \textbf{CD} and \textbf{HD}. Comparable values are reported in \cite{DiGiusto2023AnalysisHead} for full ear point clouds obtained through photogrammetry on the same dummy head with similar optical treatment, showing averaged values between the two ears of $\mathrm{\textbf{Acc}} = \SI{0.62}{\milli\meter}$, $\mathrm{\textbf{Cmp}} = \SI{93.7}{\percent}$, $\mathrm{\textbf{Avg}} = \SI{0.16}{\milli\meter}$, $\mathrm{\textbf{Max}} = \SI{3.45}{\milli\meter}$, and $\mathrm{\textbf{CD}} = \SI{0.35}{\milli\meter\squared}$. The small differences between the outcomes of the previous and current work may be attributed to the inclusion of a single scan per ear in the former study, whereas the latter includes multiple scans. While both studies utilised similar equipment and parameters, as detailed in Sec.~\ref{sec:2B}, certain factors capable of modifying the results may play a role, e.g.\ the scanning spray application, whose outcome is random in nature.

\begin{figure}[ht]
\centering
\begin{subfigure}[t]{\columnwidth}
\centering
\includegraphics[width=\columnwidth]{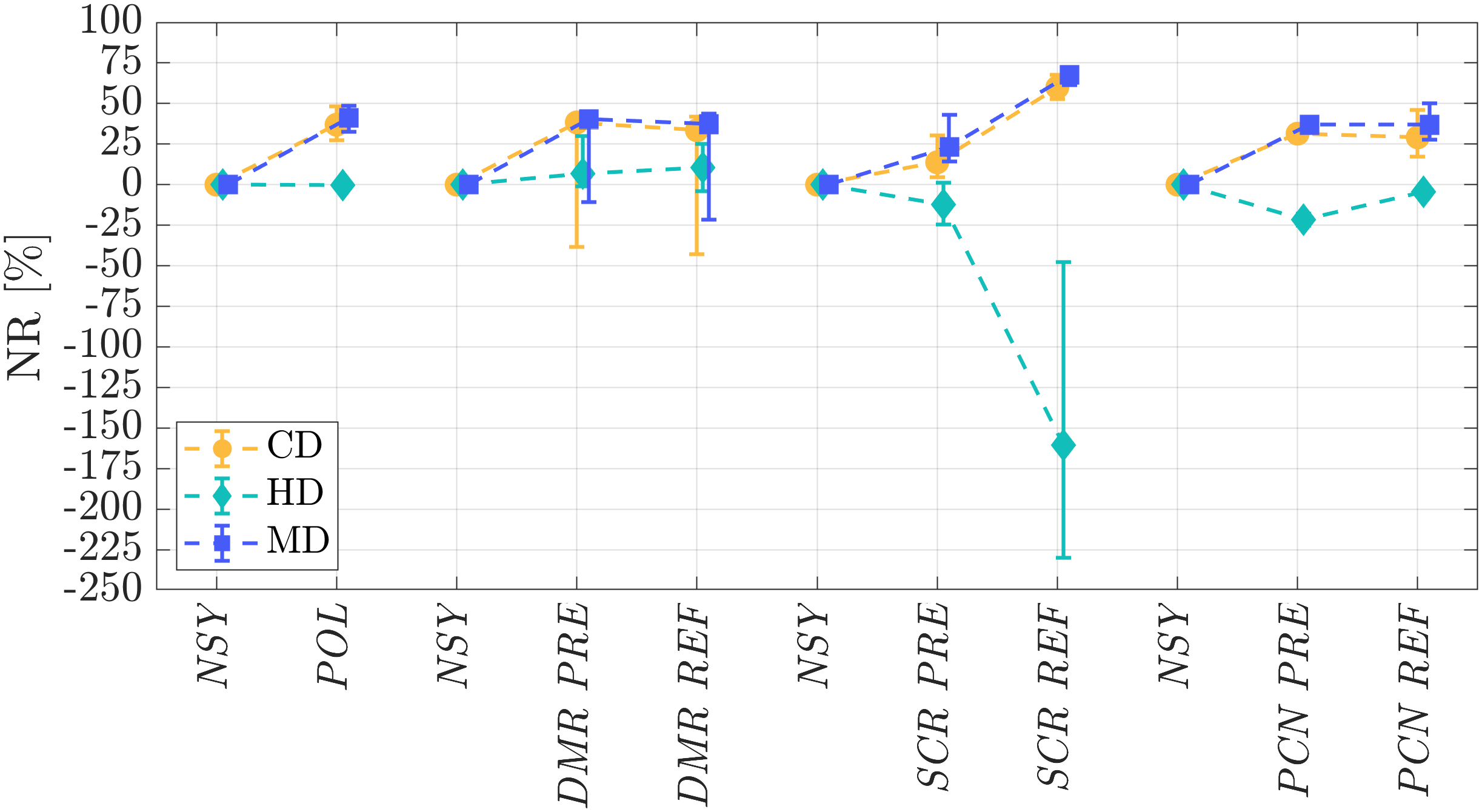}
\caption{Testingset.}
\label{fig:smr_dns_tst}
\end{subfigure}
\begin{subfigure}[t]{\columnwidth}
\centering
\includegraphics[width=\columnwidth]{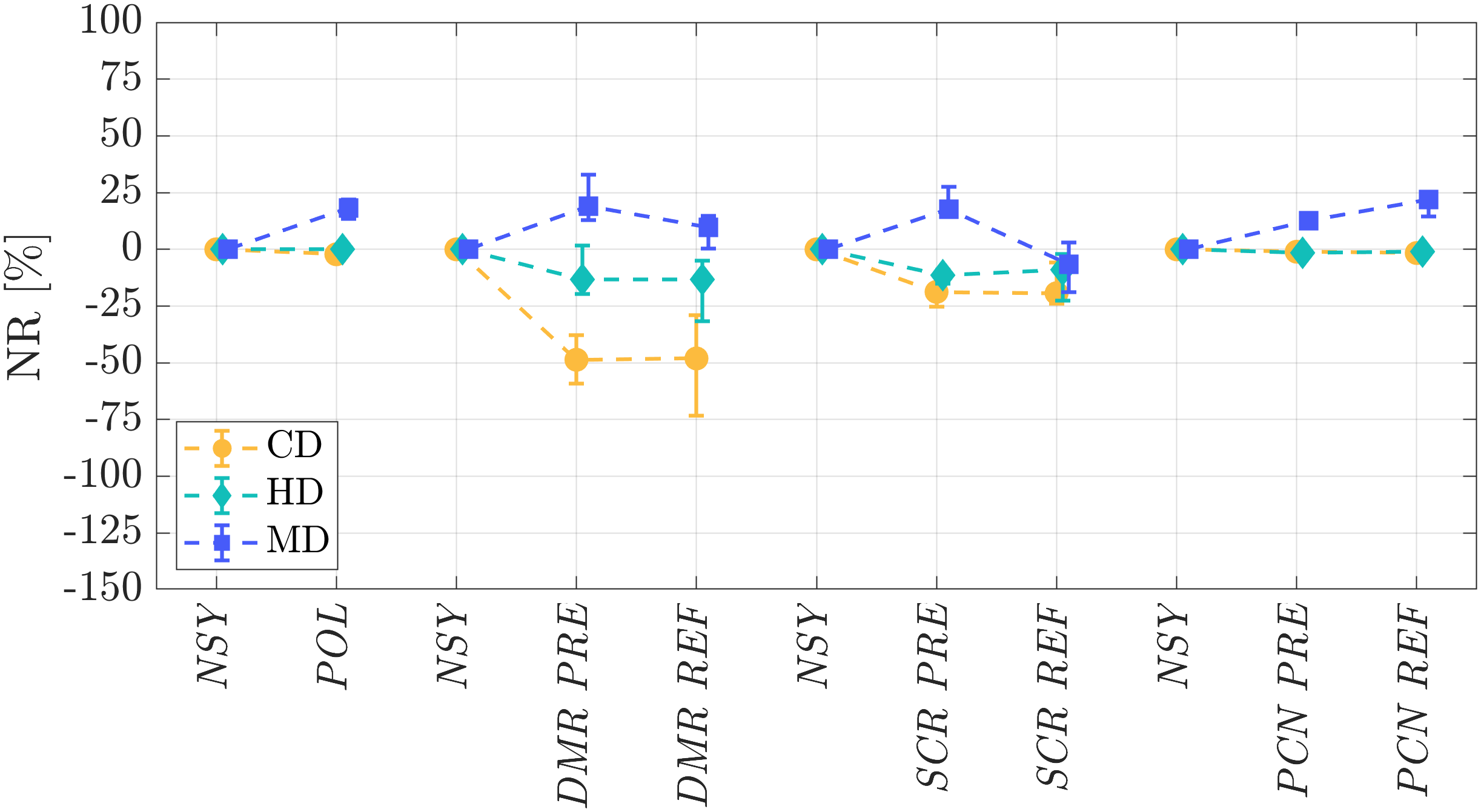}
\caption{Scanset.}
\label{fig:smr_dns_scn}
\end{subfigure}
\caption{Noise reduction obtained with the denoising algorithms on the testingset (a) and scanset (b). The bars show the interquartile range, while the markers show the median.}
\label{fig:smr_dns}
\end{figure}

The performance of the tested denoising models on the scanned ear point clouds is assessed in terms of their \textbf{NR}, calculated using Eq.~(\ref{eq:NR}). Only a subset of the distance metrics is evaluated between the tested and reference scans, i.e.\ \textbf{CD}, \textbf{HD}, and \textbf{MD}, defined in Eq.~(\ref{eq:CD}), (\ref{eq:HD}), and (\ref{eq:MD}), respectively. This choice is justified by the high correlation observed between \textbf{CD} and the geometric metrics employed in \cite{DiGiusto2023AnalysisHead}, i.e.\ all those adopted in the current study except for \textbf{HD} and \textbf{MD}. It should be mentioned that these metrics are computed on the full point clouds to avoid the manual processing steps required to extract the frontal pinna points for each tested ear shape. Various trainings of the DNNs are evaluated, including pretrained versions (\textit{PRE}), based on the original training of the models, and refined versions (\textit{REF}), fine-tuned from the pretrained weights on the ear dataset with modelled photogrammetric error. Given that the architecture of \textit{PCNm} is modified, fine-tuning starting from the pretrained weights is not feasible; therefore, the tested versions of this model are retrained from scratch on the ear dataset. The \textbf{NR} values obtained from denoising the photogrammetric point clouds with the different DNN models are compared against each other and against the outcome of \textit{POL}. The results are computed on the testing and scan datasets, referred to as testingset and scanset, containing synthetic and real photogrammetric error, respectively. The outcome is displayed in Fig.~\ref{fig:smr_dns}, showing the \textbf{NR} for the \textit{PRE} and \textit{REF} versions of the original DNNs. For clarity, the results of different models are plotted separately, starting from the initial noisy photogrammetric point cloud (\textit{NSY}), with $\mathrm{\textbf{NR}} = \SI{0}{\percent}$. The plots illustrate the distribution of \textbf{NR} across the various denoised scans by showing its interquartile range as an error bar, i.e.\ the range from the 1\textsuperscript{st} to the 3\textsuperscript{rd} quartile, and highlighting its 2\textsuperscript{nd} quartile, i.e.\ the median.

Fig.~\ref{fig:smr_dns_tst}, presenting the results on the testingset, evaluates whether model refinement through fine-tuning is beneficial, given that the point cloud errors in this set are identical to those encountered during training. It can be observed that the \textit{PRE} versions of the DNNs generally perform on par with or worse than \textit{POL}. Fine-tuning the DNNs on ear point clouds corrupted with synthetic photogrammetric error improves the performance of some algorithms. Specifically, the \textbf{NR} for \textit{SCR} shows an increase of nearly \SI{50}{\percent} in both \textbf{CD} and \textbf{MD} from \textit{PRE} to \textit{REF}, though \textbf{HD} appears to be strongly negatively impacted and exhibits high variability. Since \textbf{HD} is the only non-averaged metric, this suggests that while \textit{SCR} can reduce the overall error, it may incorrectly displace certain points. For \textit{DMR} and \textit{PCN}, fine-tuning appears to have a limited and generally negative impact on \textbf{CD} and \textbf{MD}, with only a minor positive effect on \textbf{HD}. However, for \textit{PCN}, the \textbf{NR} evaluated on \textbf{HD} remains negative in both the \textit{PRE} and \textit{REF} cases. The denoising performance of these models is also computed on the scanset, to determine whether the DNNs are effective on real scanning errors. The results, shown in Fig.~\ref{fig:smr_dns_scn}, indicate a generally lower \textbf{NR} compared to that obtained on the testingset, with all \textit{PRE} models showing comparable or worse performance than \textit{POL}. The \textit{REF} results suggest that fine-tuning tends to degrade the denoising performance of both \textit{DMR} and \textit{SCR}. For \textit{PCN}, an improvement in \textbf{MD} is observed, reaching values slightly higher than those of \textit{POL}, while \textbf{CD} and \textbf{HD} remain unaffected, displaying \textbf{NR} values close to \SI{0}{\percent} for \textit{POL} and both versions of \textit{PCN}.

\begin{figure}[ht]
\centering
\includegraphics[width=\columnwidth]{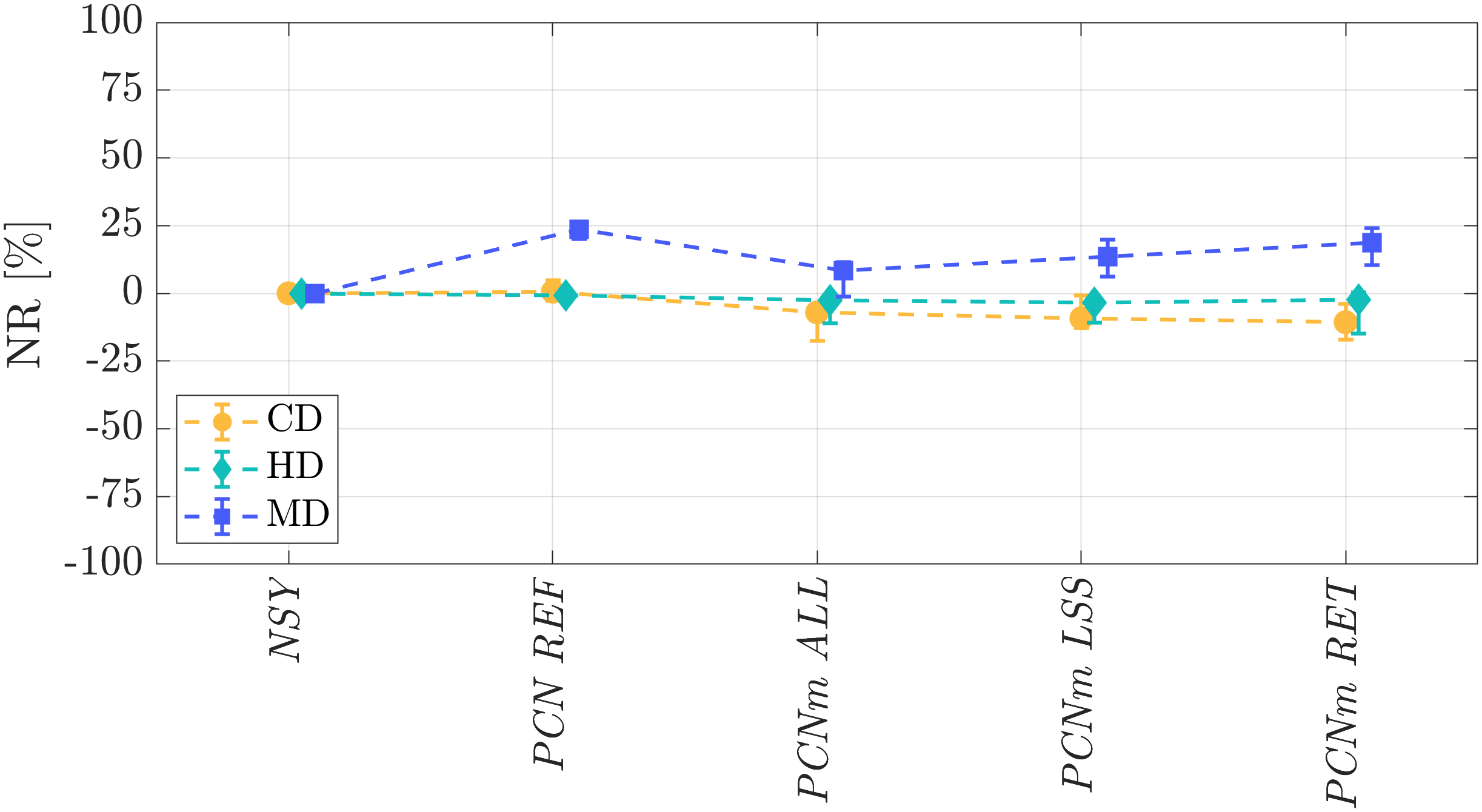}
\caption{Noise reduction obtained with different trainings of \textit{PCN} and \textit{PCNm} on the scanset. The bars show the interquartile range, while the markers show the median.}
\label{fig:smr_abl_scanset_w0}
\end{figure}

Since \textit{PCN REF} is the only model that appears to perform better than the non-DNN-based approach on the scanset, specifically in terms of \textbf{MD}, its architecture is further modified, as outlined in Sec.~\ref{sec:2C}, aiming to improve its effectiveness in reducing the error affecting photogrammetric ear scans. The modified \textit{PCN} version, referred to as \textit{PCNm}, is tested on the scanset, and the results are shown in Fig.~\ref{fig:smr_abl_scanset_w0}. Different trainings of \textit{PCNm} are evaluated and compared to \textit{PCN REF}. The first version, termed \textit{PCNm ALL}, incorporates both the modified loss function and the inclusion of the global subsample in the input data. This model is trained from scratch on the ear dataset with synthetic error. The second version, referred to as \textit{PCNm LSS}, is similar to the previous but excludes the global subsample; hence, only the loss function is modified, while the input consists solely of the local point patch. This model is also trained from scratch on the synthetic ear dataset. The last version, termed \textit{PCNm RET}, builds upon the trained \textit{PCNm LSS} by applying fine-tuning on a subset of the scanset, as described in Sec.~\ref{sec:2B}, to assess whether further refinement on samples affected by real scanning error can enhance its denoising performance. It is important to note that the \textbf{NR} for all models presented in this plot is evaluated on a subset of the scanset used for testing, derived from the original and repeated \textit{KU1} scans, which does not include any point clouds used during training or validation of \textit{PCNm RET}.

It can be seen that both modifications to this DNN lead to a deterioration in its denoising performance, as evidenced by the lower performance of \textit{PCNm ALL} in comparison to \textit{PCN REF} across all tested metrics. Although \textit{PCNm LSS} shows slightly better results, it still generally underperforms \textit{PCN REF}. Exposing \textit{PCNm LSS} to a small amount of point clouds affected by real scanning error proves beneficial, as the \textbf{NR} related to \textbf{MD} for \textit{PCNm RET} tends to be higher than that of \textit{PCNm LSS}, and in line with \textit{PCN REF}. However, when comparing the median values, \textit{PCNm RET} appears to slightly underperform \textit{PCN REF} and shows greater variability. For the \textbf{CD} and \textbf{HD} metrics, the modified DNN models generally exhibit slightly lower performance than \textit{PCN REF}.

\begin{figure}[ht]
\centering
\includegraphics[width=\columnwidth]{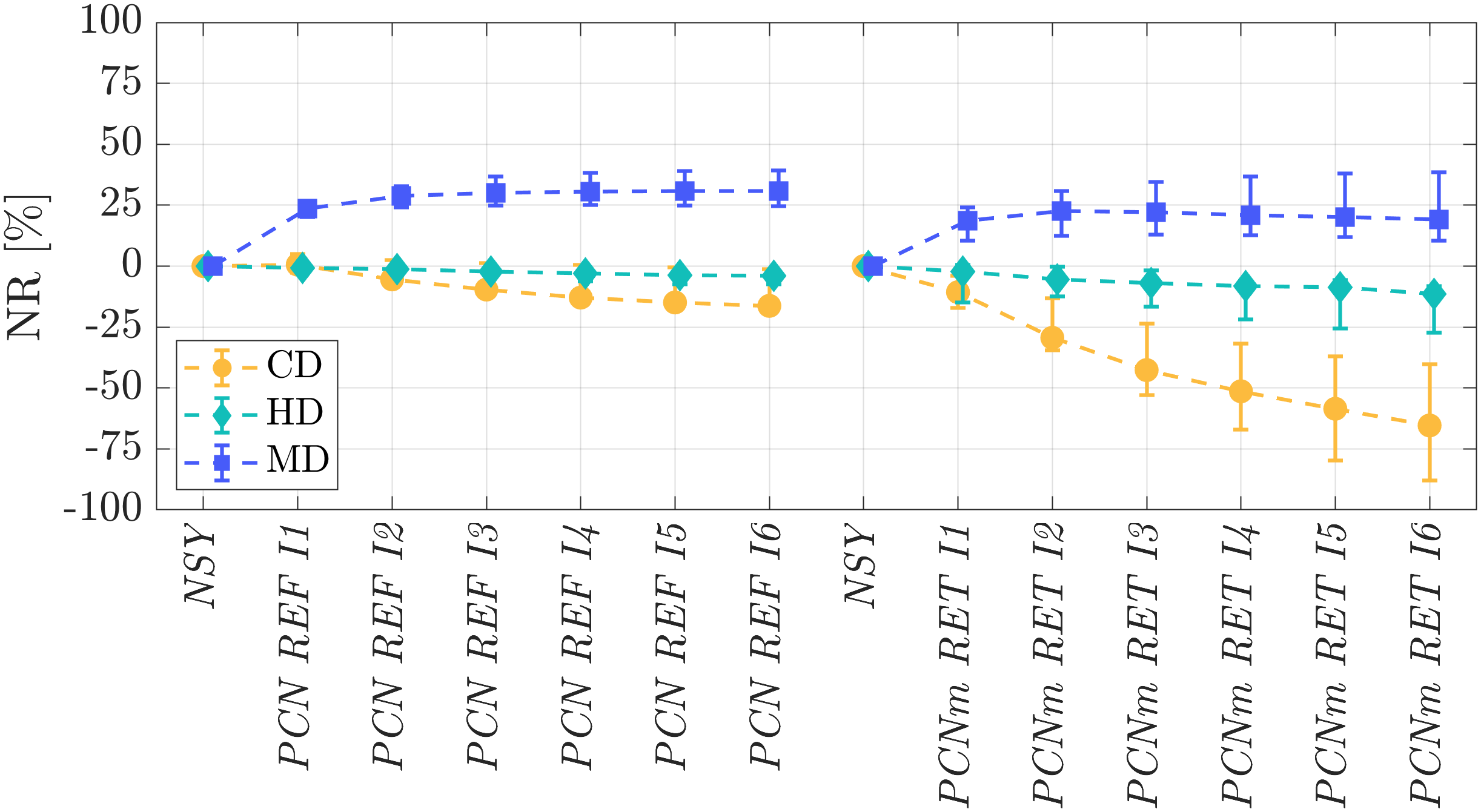}
\caption{Noise reduction obtained with the iterative application of \textit{PCN REF} and \textit{PCNm RET} on the scanset. The bars show the interquartile range, while the markers show the median.}
\label{fig:smr_dns_it}
\end{figure}

Given that \textit{PCN} can be applied iteratively to further reduce residual error in subsequent DNN iterations, this approach is tested on the scanned \textit{KU1} point clouds using \textit{PCN REF} and \textit{PCNm RET}. The resulting \textbf{NR} for the tested distance metrics is visualised in Fig.~\ref{fig:smr_dns_it}, where $6$ iterations of both models are applied, and their outcomes are compared. 

The results indicate that iterative denoising with these DNNs is beneficial for the \textbf{MD}, although reaching a plateau after a few iterations. Indeed, after the second iteration, the \textbf{NR} related to \textbf{MD} remains stable for \textit{PCN REF} and slowly decreases for \textit{PCNm RET}. Conversely, a deterioration effect is observed for \textbf{CD} and \textbf{HD} with multiple denoising passes. The \textbf{NR} for these two metrics, while generally close to \SI{0}{\percent} after the first DNN pass, shows a decline with subsequent iterations, particularly for \textit{PCNm RET}. Although the trends are similar for both \textit{PCN REF} and \textit{PCNm RET}, the latter tends to underperform the first and exhibits higher variability.
\begin{figure*}[ht]
\centering
\begin{subfigure}[t]{\columnwidth}
\centering
\includegraphics[width=\columnwidth]{Fig2_0.png}
\end{subfigure} \\
\vskip6pt
\begin{subfigure}[t]{0.48\columnwidth}
\centering
\includegraphics[width=\columnwidth]{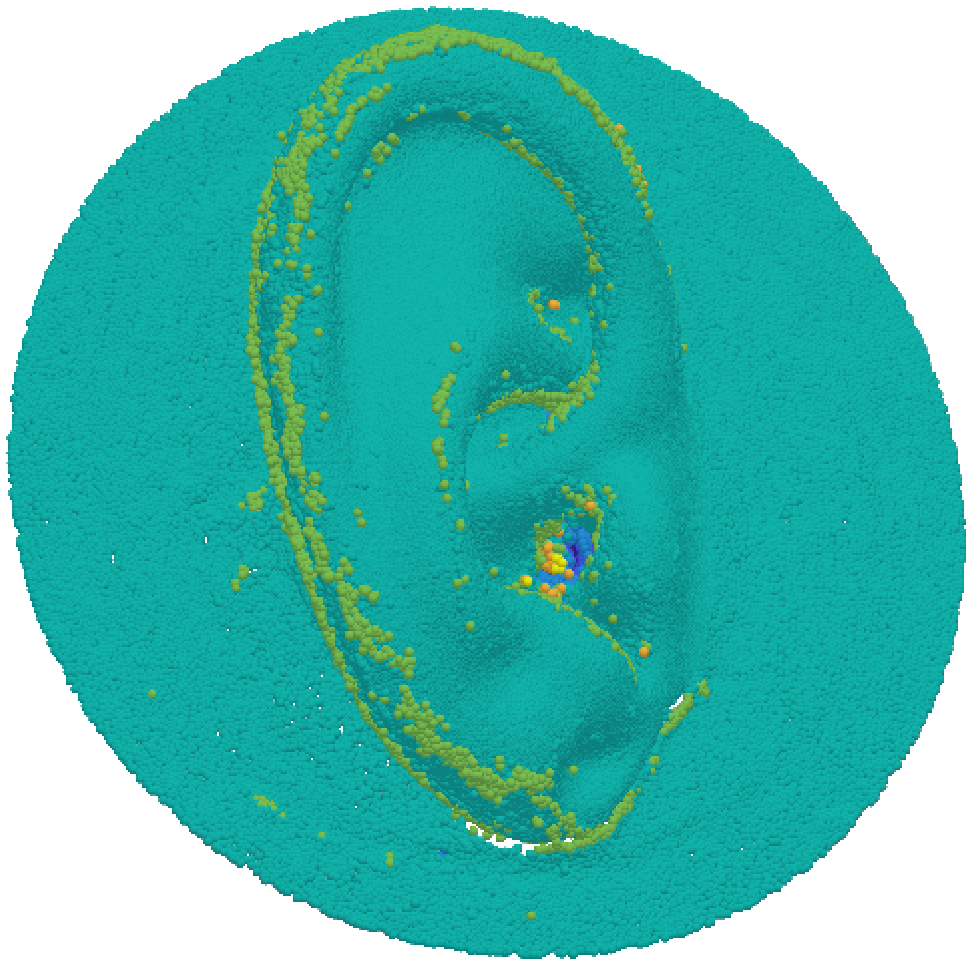}
\caption{Initial PC.}
\end{subfigure}\hfill 
\begin{subfigure}[t]{0.48\columnwidth}
\centering
\includegraphics[width=\columnwidth]{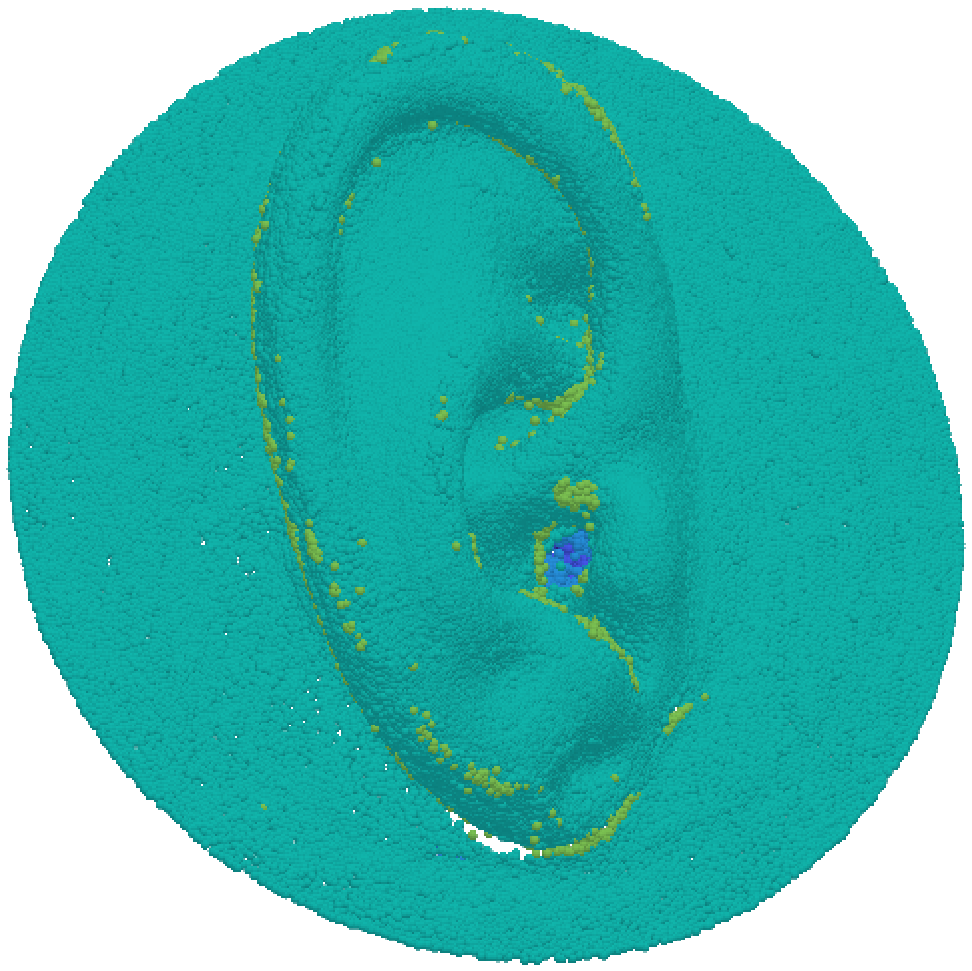}
\caption{Denoised PC (iter 1).}
\end{subfigure}\hfill 
\begin{subfigure}[t]{0.48\columnwidth}
\centering
\includegraphics[width=\columnwidth]{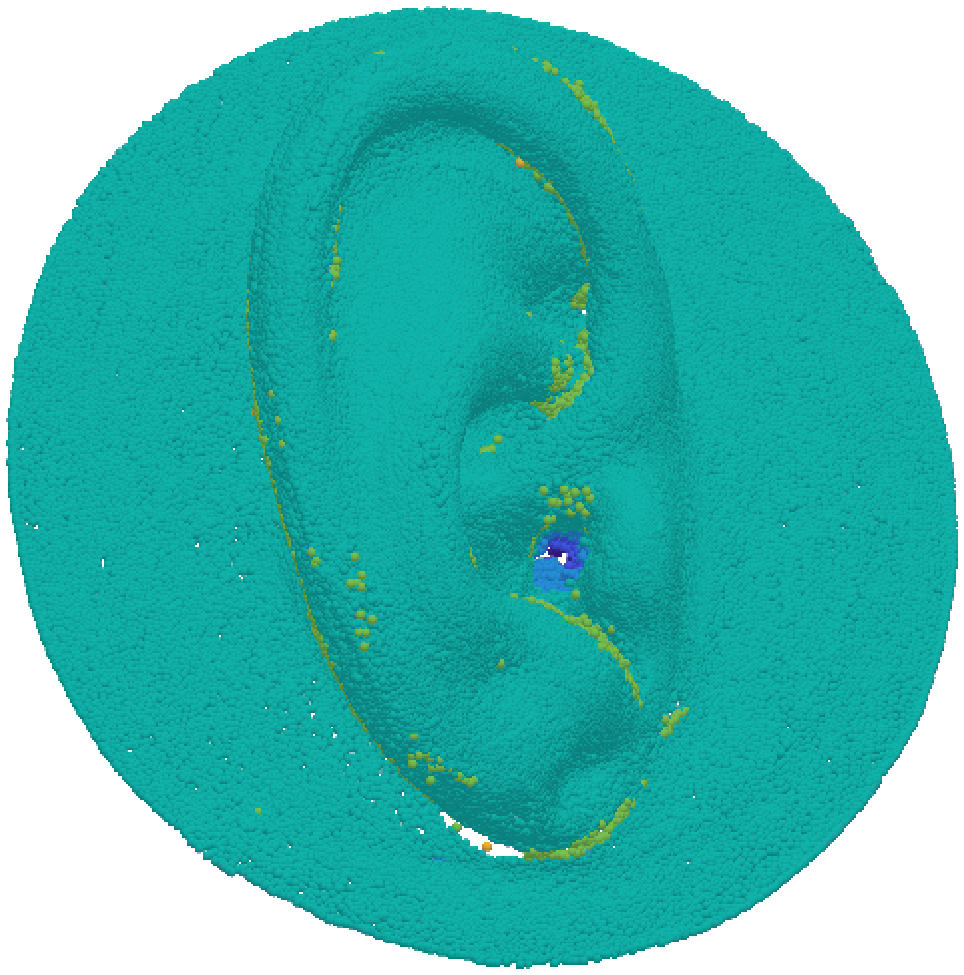}
\caption{Denoised PC (iter 3).}
\end{subfigure}\hfill 
\begin{subfigure}[t]{0.48\columnwidth}
\centering
\includegraphics[width=\columnwidth]{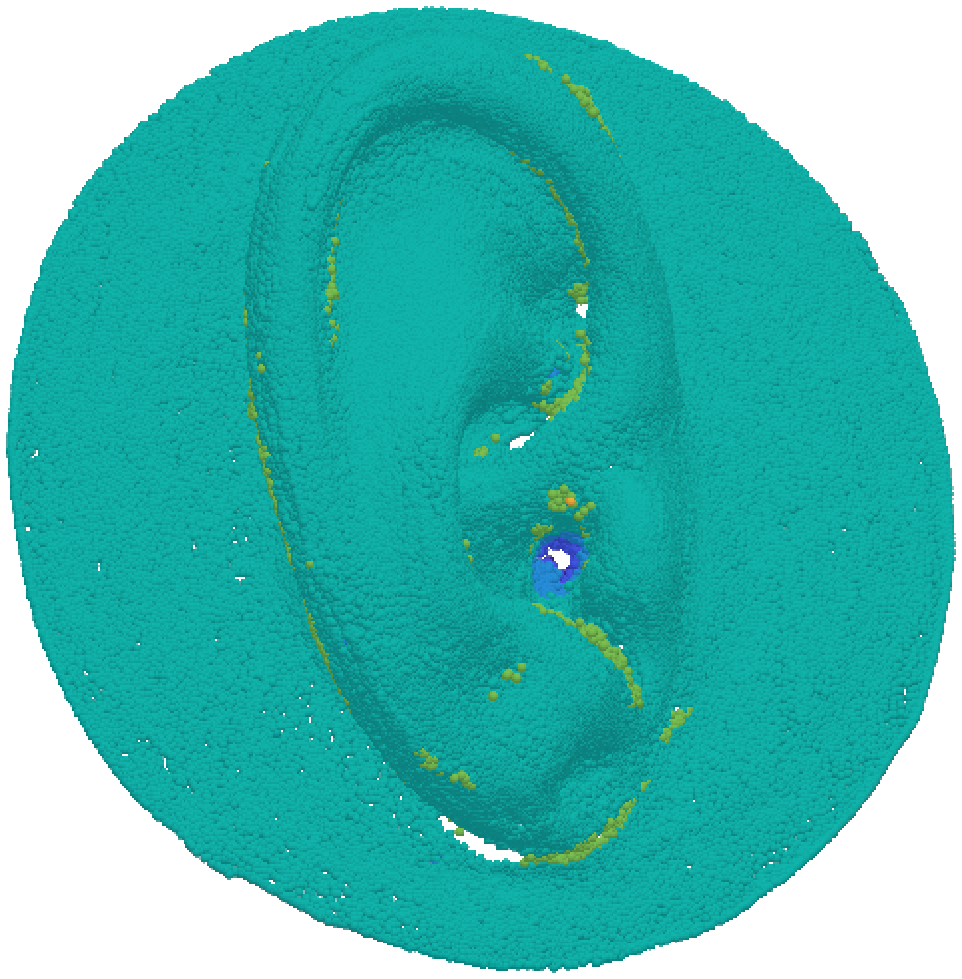}
\caption{Denoised PC (iter 6).}
\end{subfigure} \\
\begin{subfigure}[t]{0.48\columnwidth}
\centering
\includegraphics[width=\columnwidth]{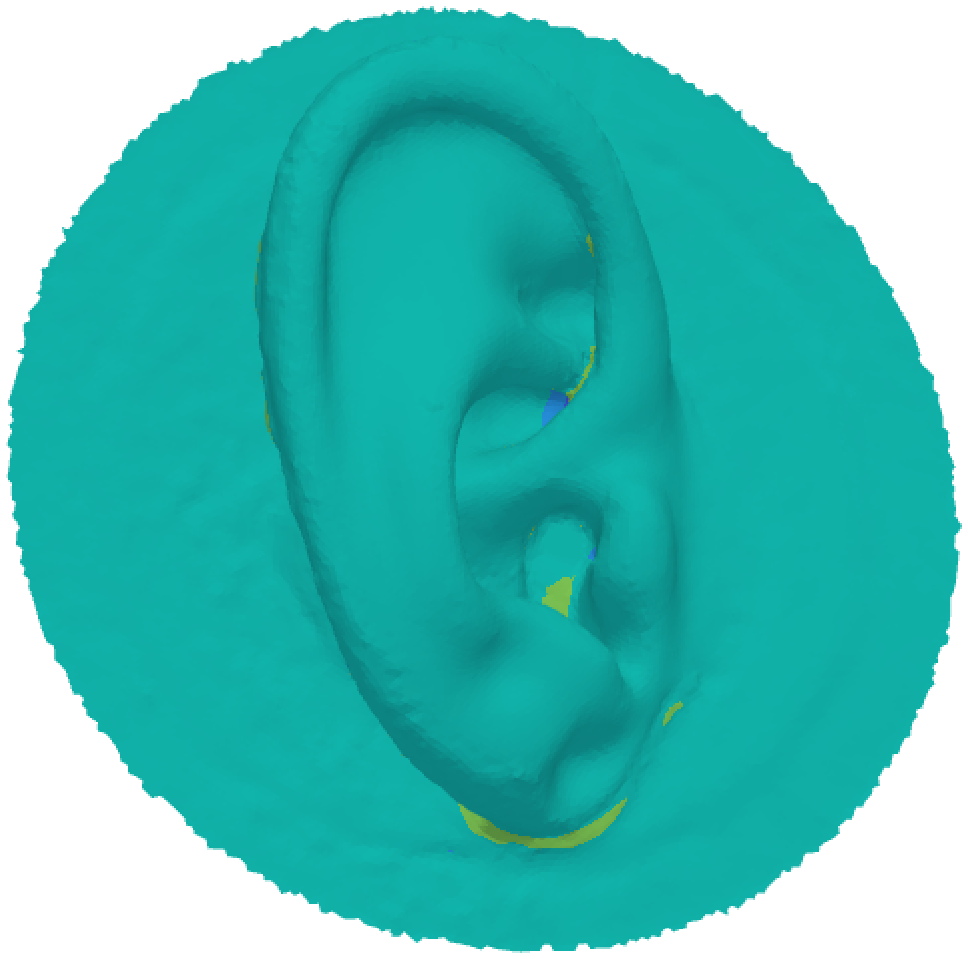}
\caption{Initial mesh.}
\end{subfigure}\hfill 
\begin{subfigure}[t]{0.48\columnwidth}
\centering
\includegraphics[width=\columnwidth]{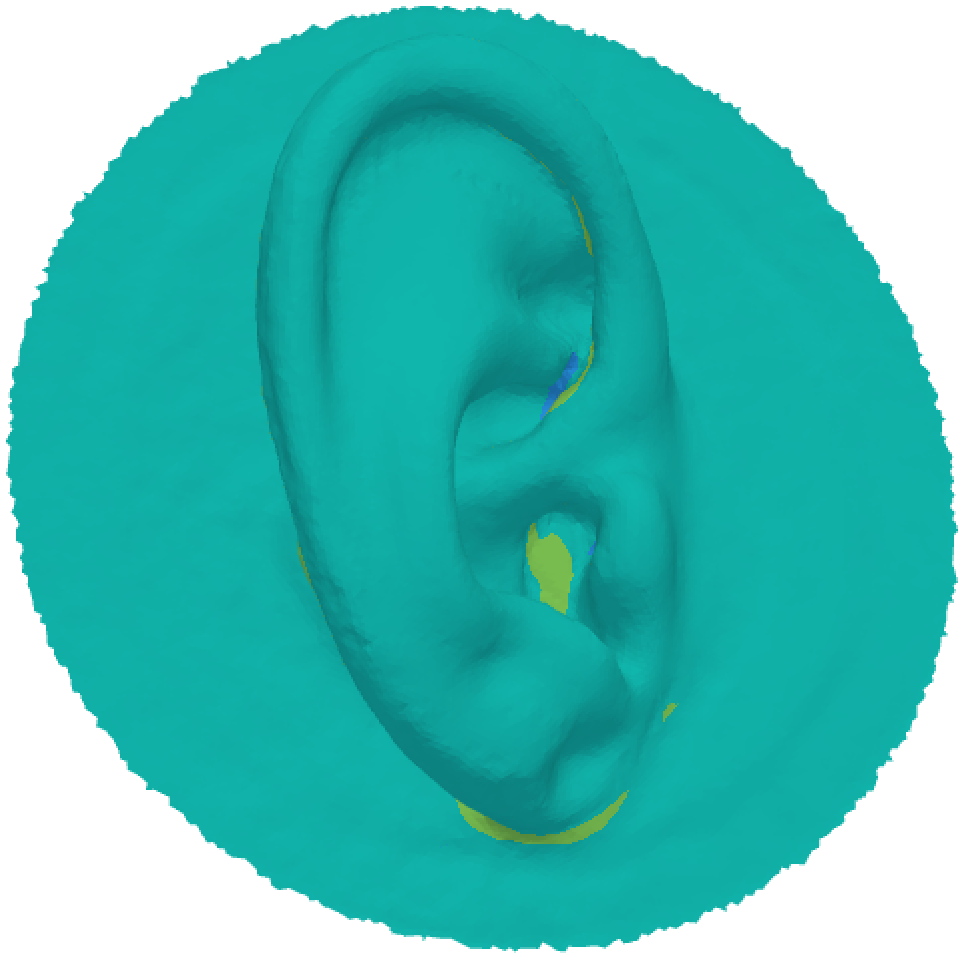}
\caption{Denoised mesh (iter 1).}
\end{subfigure}\hfill 
\begin{subfigure}[t]{0.48\columnwidth}
\centering
\includegraphics[width=\columnwidth]{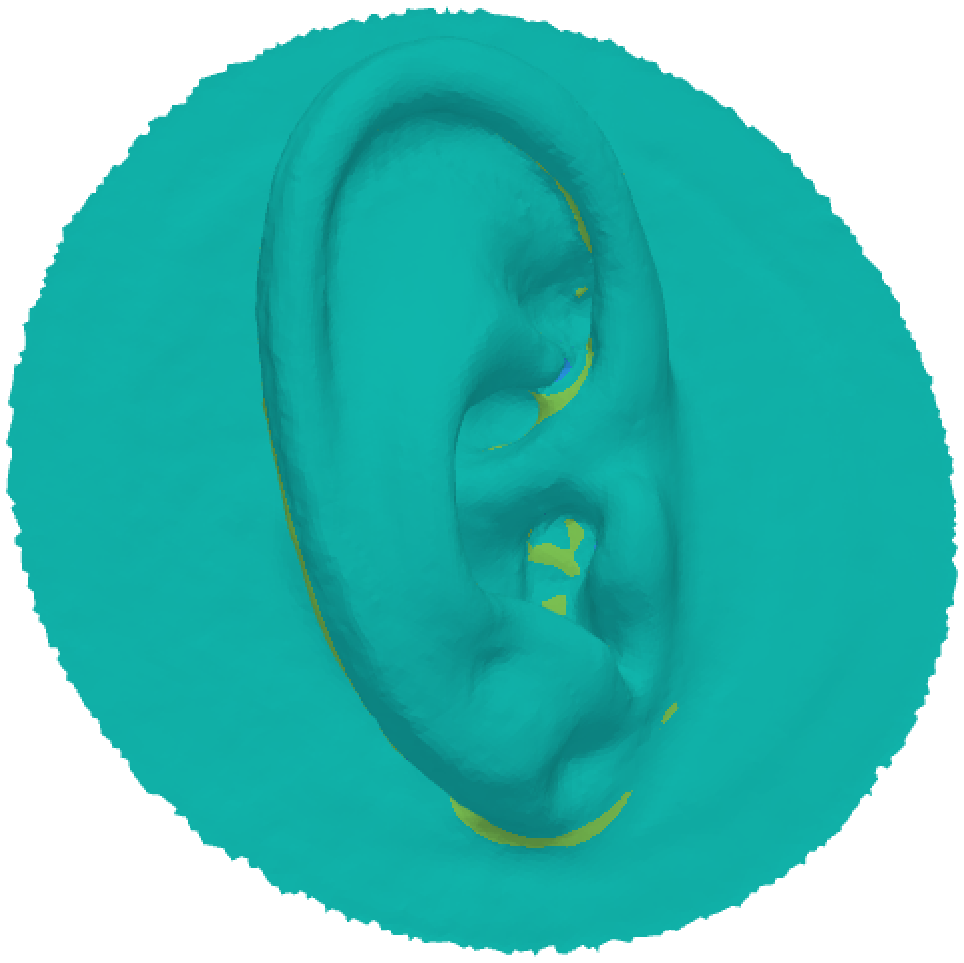}
\caption{Denoised mesh (iter 3).}
\end{subfigure}\hfill 
\begin{subfigure}[t]{0.48\columnwidth}
\centering
\includegraphics[width=\columnwidth]{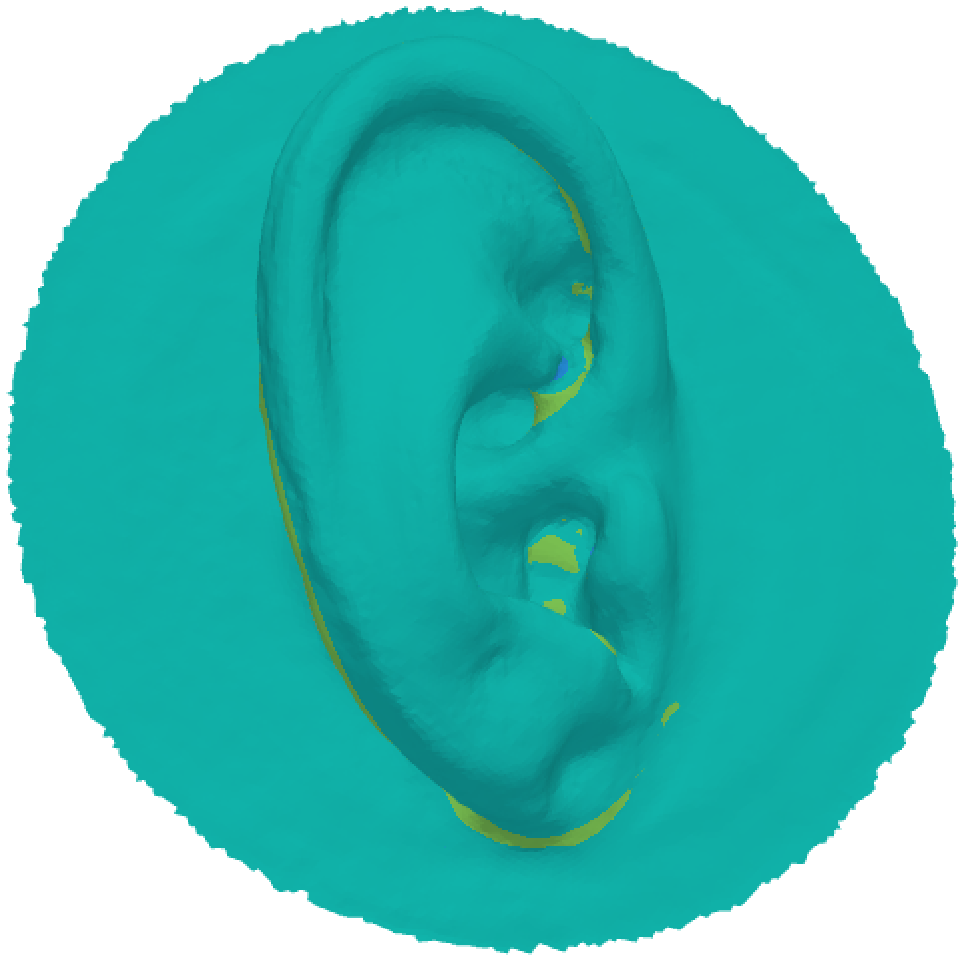}
\caption{Denoised mesh (iter 6).}
\end{subfigure}\\
\caption{\textit{KU1} right ear Point Clouds (PCs) and meshes derived from the repeated photogrammetric scan with chalk marks. Initial photogrammetric PC (a) and denoised PCs with \textit{PCNm RET} after $1$ (b), $3$ (c), and $6$ (d) iterations (iter). Meshed \textit{KU1} right ear geometries related to the initial photogrammetric PC (e) and denoised PCs with \textit{PCNm RET} after $1$ (f), $3$ (g), and $6$ (h) iterations. The colour shows the signed distance from the laser scan, cropped at $\pm\SI{2.5}{\milli\meter}$.}
\label{fig:ku1_phm2_scn_pcn_ret_iter}
\end{figure*}

A visualisation of the iterative denoising effect with \textit{PCNm RET} on a photogrammetric scan is presented in Fig.~\ref{fig:ku1_phm2_scn_pcn_ret_iter}, where the point cloud derived from the repeated right ear \textit{KU1} scan with chalk marks is displayed in its initial stage, i.e.\ the raw photogrammetric output, and after a single or multiple denoising iterations. Furthermore, the mesh corresponding to each point cloud is also shown to highlight the effect of the meshing procedure in the presence of different error levels. The SDF between these ear shapes and the \textit{KU1} laser mesh is overlaid onto each geometry.

Focusing on the point clouds, it is evident that \textit{PCNm} is capable of effectively reducing the overall amount of error in this scan. The noisy points appearing at the outer part of the helix are almost entirely displaced towards the underlying surface after the first iteration, while some error appears to persist at the most concave pinna locations, e.g.\ the cavum and cymba conchae, even after multiple DNN passes. It is also noticeable that, after several denoising iterations, holes tend to emerge in the most occluded pinna structures. These areas of the photogrammetric scan generally exhibit few points and high error due to limited visibility. The drop in \textbf{CD} and \textbf{HD} with multiple denoising iterations, depicted in Fig.~\ref{fig:smr_dns_it}, is thought to arise from the emergence of holes in the scanned point clouds. These metrics, taking into account the distance from reference to tested point cloud, are highly sensitive to such anomalies. The meshes related to each displayed point cloud show that the adopted meshing algorithm can cope with some of the photogrammetric error. This is noticeable for the mesh obtained on the initial photogrammetric scan, where the high level of error affecting certain pinna locations does not seem to impact the related mesh. A similar situation is observed for the point cloud obtained after the first denoising iteration. Conversely, with further iterations, the error tends to concentrate at specific ear locations, e.g.\ the cymba, and the meshing algorithm appears to interpolate between the noisy points, creating artefacts in the meshed ear surface. Moreover, the holes that emerge after multiple denoising passes further degrade the meshed geometry. This is attributed to the meshing and hole-filling algorithms, which interpolate between the denoised points and artificially fill some of the concave pinna structures, crucial for the creation of the HRTF spectral features \cite{DiGiusto2023AnalysisHead}.

\subsection{HRTFs}
\label{sec:3B}
Given that the relation between ear morphology and HRTFs is not yet fully understood \cite{Pollack2022PerspectiveOverview}, further analyses are necessary to evaluate the denoising effect on the HRTFs simulated on meshes originating from the processed point cloud data. Therefore, the HRTFs of the initial and denoised photogrammetric \textit{KU1} scans are computed and compared to those related to the laser-scanned data. For this dummy head, repeated HRTF measurements are available, and a method is proposed in \cite{DiGiusto2023AnalysisHead} to select and group these HRTFs based on their similarity to a reference set computed on a laser-scanned geometry. These results are employed in the current study to better assess the outcomes obtained on several denoised scans of the same dummy head. The procedure defined in Sec.~\ref{sec:2D} is leveraged to obtain meshes from the initial photogrammetric scans and processed point clouds with different denoising algorithms. To compare the results of DNNs with similar training, the HRTFs are computed on point clouds denoised with their \textit{REF} versions. Furthermore, \textit{PCNm RET} is also tested. For both \textit{PCN REF} and \textit{PCNm RET}, the denoised point clouds after a single and multiple iterations are analysed to assess the effect of the iterative approach on the HRTFs. As done in a previous study \cite{DiGiusto2023AnalysisHead}, frequency scaling is applied to the numerical HRIRs using the value of $\alpha = 0.99$, obtained by averaging the optimal scaling factors evaluated between the laser and measured \textit{KU1} DTFs, relating to a mean ISSD decrease of \SI{0.06}{\deci\bel\squared}.

The median ISSD, computed with Eq.~(\ref{eq:ISSD}) between the laser scan DTFs and those obtained on the initial and denoised \textit{KU1} photogrammetric scans, is presented in Tab.~\ref{tab:issd_qe_pe}. Moreover, this table includes the median QE and PE deviation obtained by applying the sagittal localisation model with template laser-scanned DTFs and target initial photogrammetric and denoised point clouds DTFs. The reference error values relate to a baseline condition in which the laser-scanned DTFs are used as both the template and the target, corresponding to localisation performance with individual auditory filters \cite{Baumgartner2014ModelingListeners}. Since the photogrammetric error could impact the left and right ear differently, and to further increase the amount of available data, reference and tested left ear DTFs are mirrored to the right, and vice versa. Thus, the median deviation is computed across all the mirrored data, using the related mirrored version of the laser scan DTFs as reference. Therefore, the median value in each metric is evaluated across $8$ samples derived from the left and right mirrored DTFs obtained on the $4$ \textit{KU1} scans, comprising $2$ original and $2$ repeated scans.
\begin{table}[ht]
\centering
\caption{Median deviation from reference of objective and perceptual metrics evaluated across the mirrored left and right ear DTFs derived from the scanned and denoised \textit{KU1} point clouds. The reference DTF is derived from the laser scan. The highlighted results show to the lowest median deviation in each metric.}
\label{tab:issd_qe_pe}
\vskip3pt
\begin{tabular}{cccc}
\hline\hline
ID & \multicolumn{3}{c}{Median deviation from reference} \\ \cline{2-4}
 &ISSD $[\SI{}{\deci\bel\squared}]$ &QE $[\SI{}{\percent}]$ &PE $[\SI{}{\degree}]$\\
\hline
\textit{PHO} &1.54 &8.53 &4.83\\
\textit{POL} &2.05 &9.81 &5.70\\
\textit{DMR REF} &1.58 &8.48 &3.96\\
\textit{SCR REF} &1.88 &10.41 &6.15\\
\textit{PCN REF I1} &1.50 &8.71 &4.42\\
\textit{PCN REF I3} &1.86 &10.91 &6.04\\
\textit{PCN REF I6} &1.84 &11.86 &6.31\\
\textit{PCNm RET I1} &\textbf{1.28} &\textbf{5.58} &\textbf{3.21}\\
\textit{PCNm RET I3} &2.19 &9.00 &4.22\\
\textit{PCNm RET I6} &2.31 &10.30 &6.73\\
\hline\hline
\end{tabular}
\end{table}

The results indicate a relatively low deviation in the initial photogrammetric case (\textit{PHO}), especially in comparison to a previous study where the DTFs derived from a single photogrammetric scan of the same dummy head were tested against a similarly obtained reference. The reported deviations in ISSD, QE, and PE were \SI{4.5}{\deci\bel\squared}, \SI{16.4}{\percent}, and \SI{12.0}{\degree}, respectively \cite{DiGiusto2023AnalysisHead}. The lower divergence in the current study may stem from better scanning conditions or higher operator expertise, particularly in the repeated scans. It is evident that only a limited set of denoising methods appears to be beneficial for the DTFs. Specifically, only \textit{DMR}, \textit{PCN REF I1}, and \textit{PCNm RET I1} exhibit median deviations smaller or comparable to those observed for \textit{PHO}. Notably, \textit{PCNm RET I1} consistently yields the best results across all tested metrics. However, further denoising iterations with \textit{PCNm} seem to have a detrimental effect, as the deviation steadily increases from the first iteration onward, peaking with \textit{PCNm RET I6}. A similar trend is observed for multiple iterations of \textit{PCN REF}. This phenomenon is thought to relate to the effect depicted in Fig.~\ref{fig:ku1_phm2_scn_pcn_ret_iter}, where artefacts in the ear meshes tend to emerge at the concave pinna structures after successive denoising passes.

\begin{figure}[ht]
\centering
\includegraphics[width=\columnwidth]{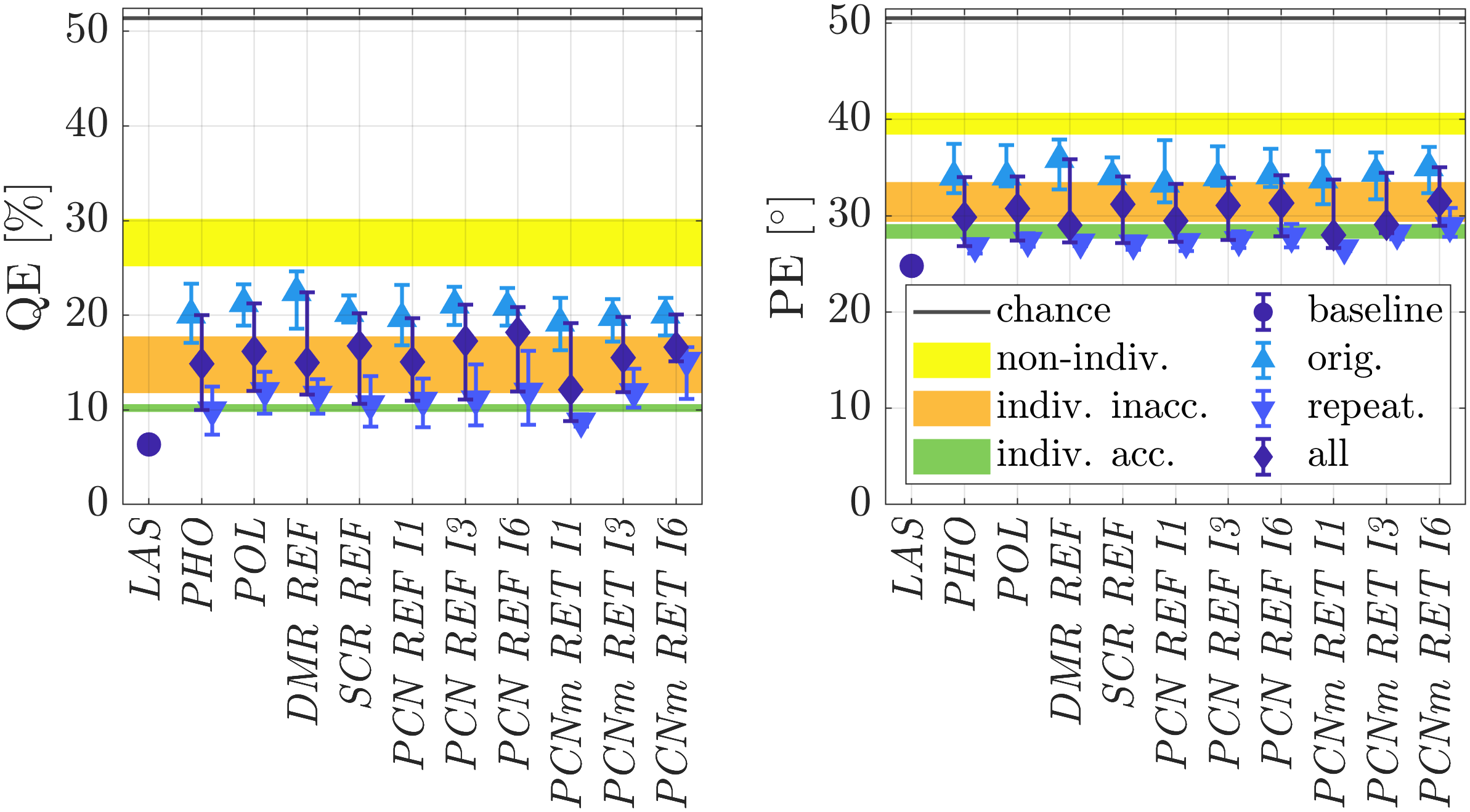}
\caption{Sagittal plane localisation error with template laser \textit{KU1} DTFs and different scanned and denoised target DTFs. The bars show the interquartile range, while the markers show the median evaluated across the mirrored left and right ear DTFs derived from the original, repeated, and all \textit{KU1} scanned and denoised point clouds. The envelopes show the interquartile range of the errors obtained with several individual and non-individual measured DTFs. The black line corresponds to chance rate, i.e. random guessing.}
\label{fig:stats_qe_pe}
\end{figure}

The results obtained on the QE and PE metrics for the original and denoised photogrammetric scans are also plotted in Fig.~\ref{fig:stats_qe_pe}, comparing them with the expected ranges of localisation error obtained using \textit{KU1} DTFs and non-individual filters. The data used to obtain the error distributions with individual and non-individual DTFs is the same as in \cite{DiGiusto2023AnalysisHead}. However, a slight difference is present, as the error ranges are derived from the distributions of localisation error with mirrored versions of the DTFs for both left and right ears, rather than conventional unmirrored DTFs, in line with how the DTFs are processed in the current study. The results are presented as the distribution obtained across all \textit{KU1} scans, as well as separately for the original and repeated scans.

Given that the QE and PE metrics are highly correlated \cite{DiGiusto2023AnalysisHead, Stitt2021SensitivityModel}, only the QE results are discussed in detail. Nonetheless, the trends of these two error metrics are similar. The outcome indicates that the median value for \textit{PHO}, computed across all scans, falls within the individual inaccurate group, related to DTFs showing limited differences in localisation performance from the reference. However, high variability is observed, with the 1\textsuperscript{st} quartile being in line with individual accurate DTFs, and the 3\textsuperscript{rd} being outside the individual group, yet lower than non-individual data. While the majority of the models exhibit worse outcomes than \textit{PHO}, some show improvements in the median value, though their interquartile range is generally in line with or worse than that of \textit{PHO}. The outcome of \textit{PCNm RET I1} displays an overall median value and interquartile range indicating lower errors than \textit{PHO}. Analysing the original and repeated scans separately reveals that the trends in localisation error are consistent with the results from all scans, with \textit{PCNm RET I1} showing the best outcome. However, the original scans start with high deviation in the \textit{PHO} case, generally above the individual inaccurate set. The denoising methods show limited effectiveness in this case, as the best outcome still results in median errors higher than those of the individual inaccurate group. Conversely, the results of the repeated scans, exhibiting less geometric deviation from the reference, show smaller initial localisation errors, falling between the individual accurate and inaccurate ranges. Also for these scans, denoising with \textit{PCNm RET I1} proves beneficial, as the median deviation of the HRTFs computed on them is below the individual accurate range, approaching the baseline localisation error. However, further iterations of this denoising algorithm appear detrimental, likely due to the effects observed in the meshes related to the iterative denoising with this DNN.

\subsection{Correlation}
\label{sec:3C}
Knowledge of the correlation between the geometric metrics evaluated on an ear point cloud and the HRTF metrics computed on its related mesh is beneficial. This insight can serve as a guide in selecting a relevant loss function for training the denoising algorithms. Therefore, the Pearson's correlation coefficient between the geometric point cloud metrics and all the employed HRTF metrics is calculated. Initially, the correlation between the HRTF metrics alone is analysed. The outcome indicates coefficients of \SI{0.83}{} between ISSD and QE, \SI{0.90}{} between ISSD and PE, and \SI{0.95}{} between PE and QE. These values are in line with those reported in \cite{DiGiusto2023AnalysisHead}, although slightly lower. This is attributed to the higher number of samples used in the current analysis compared to the limited amount of observations in the latter study.

\begin{figure}[ht]
\centering
\includegraphics[width=\columnwidth]{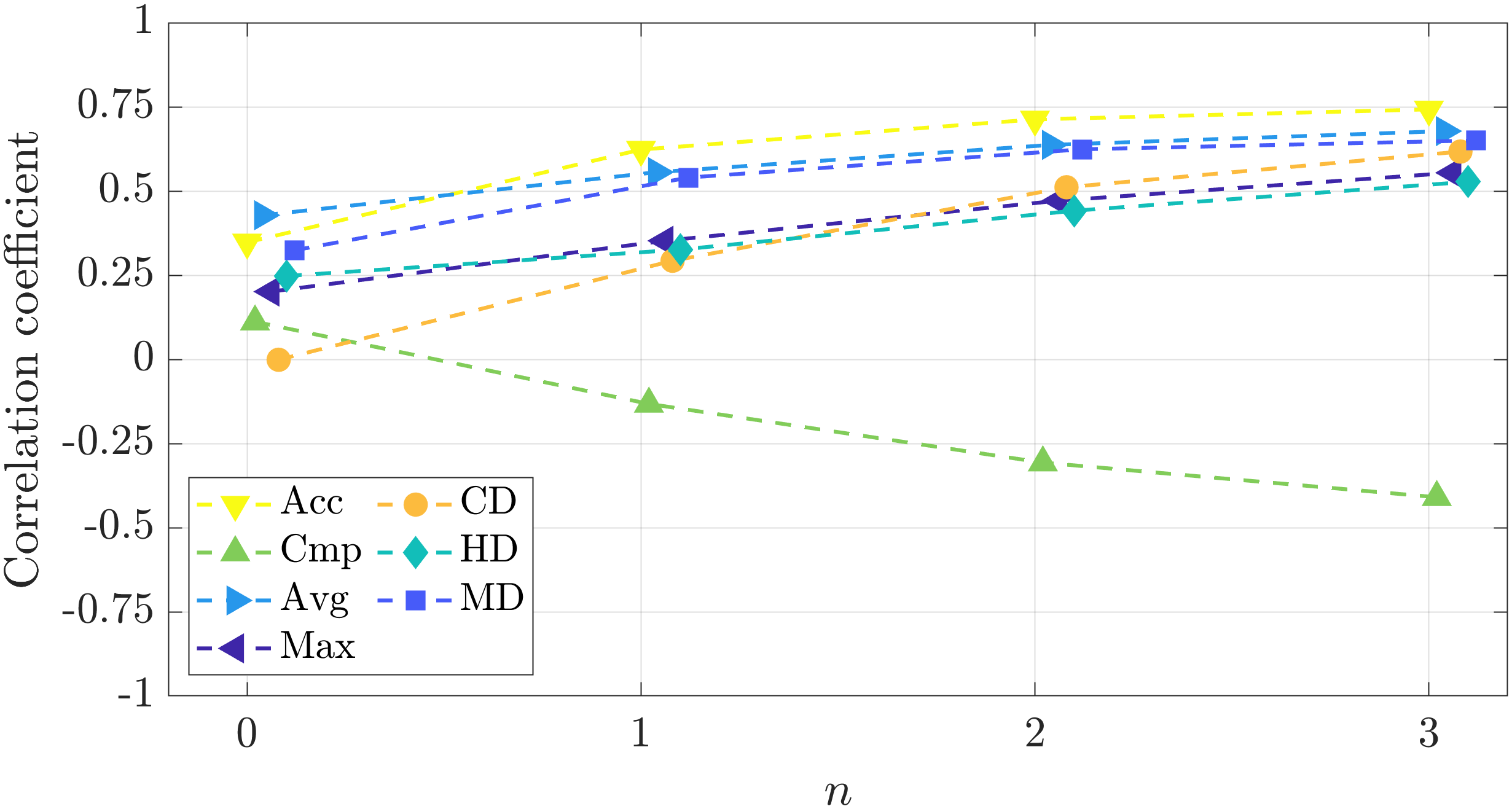}
\caption{Median correlation coefficients between geometric and HRTF metrics computed on the full scanned and denoised \textit{KU1} ear point clouds and related HRTFs, as a function of the power used to raise the AO factor for distance weighting.}
\label{fig:corr}
\end{figure}

The correlation coefficients between each of the geometric metrics computed on the full ear point clouds, including the back of the ear, and all the tested HRTF metrics are illustrated in Fig.~\ref{fig:corr}. The displayed median values are evaluated among the correlation coefficients of ISSD, QE, and PE with the geometric metrics. Moreover, these results highlight the influence of the distance weighting, since the results are plotted as a function of the integer power ($n$) used to elevate $\widehat{\mathrm{AO}^c}$ for weighting the distance between point clouds and reference geometry prior to computing the assessment metrics.

The results indicate that the correlation between the acoustic and all the geometric metrics is relatively low when no weighting is applied, i.e.\ for $n = 0$. In this case, the maximum correlation, seen for \textbf{Avg}, does not exceed \SI{0.43}{}, while several other metrics show values close to \SI{0}{}, e.g.\ \textbf{CD} and \textbf{Cmp}. The correlation tends to steadily increase with $n$, suggesting that the distance weighting gives more importance to points at acoustically relevant pinna locations. The majority of the metrics reach coefficients above \SI{0.5}{} at $n = 3$. Only \textbf{Cmp} exhibits a lower outcome, maximally reaching an absolute value below \SI{0.41}{} for the highest $n$. The negative coefficient observed for this metric is due to the fact that, in contrast to all others, a higher value of \textbf{Cmp} relates to a better scanning outcome. Consequently, there is an inverse relationship between this and the HRTF metrics. The \textbf{Acc} shows the best correlation in the weighted distance for all tested $n$ values.

\begin{figure}[ht]
\centering
\includegraphics[width=\columnwidth]{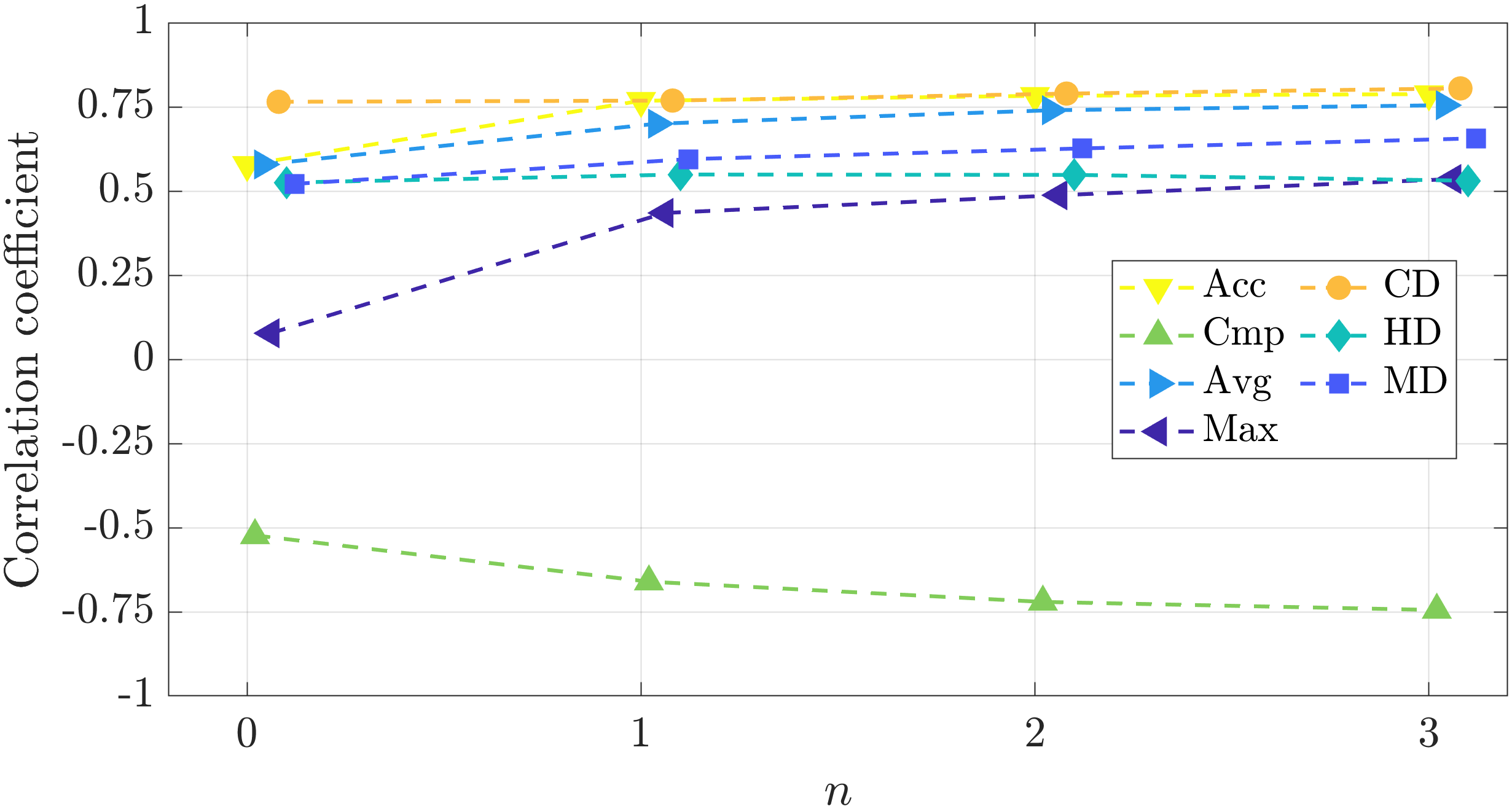}
\caption{Median correlation coefficients between geometric and HRTF metrics computed on the front part of the scanned and denoised \textit{KU1} ear point clouds and related HRTFs, as a function of the power used to raise the AO factor for distance weighting.}
\label{fig:corr_f}
\end{figure}

The same analysis is repeated, focusing solely on the frontal part of the ear, rather than also including points from the back. The extraction of the frontal pinna geometry from the full point clouds is achieved as defined in Sec.~\ref{sec:3A}. The geometric metrics are evaluated on the weighted and unweighted distances computed on these points, and their outcomes are displayed in Fig.~\ref{fig:corr_f}.

The results observed in this plot, while showing similar trends to those seen in Fig.~\ref{fig:corr}, reveal higher correlation coefficients. For the unweighted case, the correlation is generally above \SI{0.5}{}. The value close to \SI{0}{} observed for the unweighted \textbf{Max} is considered to be an artefact, since this metric is used to extract points in the frontal ear part; hence, all the point clouds have a common value of $\textbf{Max} = \SI{2}{\milli\meter}$ for $n = 0$. The highest value of \SI{0.76}{} is obtained for \textbf{CD}, while \textbf{Cmp} shows a negative correlation coefficient of \SI{-0.52}{}. A similar analysis conducted in \cite{DiGiusto2023AnalysisHead} shows higher correlation reaching absolute values close to \SI{0.84}{} for \textbf{CD} and \textbf{Cmp}, while being lower but above \SI{0.64}{} for the other metrics. The inferior coefficients obtained in the current analysis could be due to the higher number of samples, i.e.\ \SI{40}{} left and right ears, compared to the limited set of \SI{7}{} observations in the previous study. The correlation tends to rise with the factor used to weight the distance, although this increase is less steep than in the results relating to the full ear point clouds. The highest correlations are observed for $n = 3$, with \textbf{CD} reaching a maximal value of \SI{0.80}{}, while \textbf{Acc} and \textbf{Avg} show slightly lower results, still above \SI{0.75}{}. \textbf{Cmp} shows comparable but negative correlation coefficients, reaching \SI{-0.74}{} for $n = 3$.

\section{Discussion}
\label{sec:4}
The findings presented in Sec.~\ref{sec:3A} indicate that the original denoising models effectively reduce some of the geometric errors affecting the photogrammetric ear scans, particularly regarding the \textbf{MD} metric. However, the performance appears to depend on the specific model employed and the chosen assessment metric. Indeed, when examining \textbf{CD} and \textbf{HD}, denoising often proves ineffective or yields even poorer results. Moreover, the DL-based methods tend to underperform the classical denoising approach, likely due to differences in shapes and noise type present in the original training datasets. Directly comparing the results of the tested DNNs on their original data to the outcomes on the ear dataset is challenging due to variations in shape, noise, and metrics used for the assessment. Nonetheless, in \cite{Luo2021Score-BasedDenoising}, all these models were applied to denoise a common dataset of $20$ generic shapes perturbed by Gaussian noise, and mainly evaluated using \textbf{CD}. While this study examined various sampling resolutions and noise levels, the closest case to the current analysis relates to point clouds of \SI{50000}{} points corrupted by Gaussian noise with a standard deviation set to \SI{1}{\percent} of the bounding sphere's radius. The \textbf{CD} calculated for this data corresponds to a \textbf{NR} close to \SI{62}{\percent} for \textit{SCR}, and lower values around \SI{40}{\percent} for \textit{PCN} and \textit{DMR}. Applying the \textit{PRE} versions of these models to the testingset, lower \textbf{NR} values for \textbf{CD} are observed, between \SIlist{13; 34}{\percent}, while on the scanset the outcome is considerably worse. This discrepancy can be attributed to the differences in data compared to that used for training, on which the DNNs struggle to generalise.

Leveraging fine-tuning to address this issue proves beneficial for certain models, e.g.\ \textit{SCR REF} achieves $\mathrm{\textbf{NR} = \SI{60}{\percent}}$ for \textbf{CD} on the testingset; however, it does not seem to benefit the other DNNs. Furthermore, the improved performance of \textit{SCR} on the testingset does not translate to a reduction in real scanning error, as the \textbf{CD} metric is not improved by refinement. This also appears to be the case for the \textbf{MD} metric, where a \textbf{NR} around \SI{69}{\percent} was observed when applying \textit{SCR} to its original dataset \cite{Luo2021Score-BasedDenoising}, and denoising on the testingset with its \textit{PRE} and \textit{REF} versions results in \textbf{NR} values of \SIlist{23; 68}{\percent}, respectively. Notably, the performance of \textit{SCR PRE} and \textit{REF} is even lower on the scanset, around \SIlist{18; -7}{\percent}, respectively. The reasons for the differing behaviours of the DNNs depending on the input data are challenging to clearly define, and may relate to the specific architecture and denoising strategy of each model. Fine-tuning \textit{PCN} on the ear dataset enhances its performance, particularly regarding the \textbf{MD} metric. However, the improvement is limited, as the \textbf{NR} obtained by applying its \textit{REF} version on the testingset exhibits a median around \SI{37}{\percent}, in line with the \textit{PRE} model, though the interquartile range reaches higher values close to \SI{50}{\percent}. On the scanset, an increase in the median \textbf{NR} of approximately \SI{9}{\percent} is observed. This suggests that refining the DNN on data with errors more closely aligned to the target can yield better performance.

The modifications applied to \textit{PCNm}, after training it from scratch on the ear dataset, do not seem to improve its denoising performance in comparison to the fine-tuned original model, as observed in Fig.~\ref{fig:smr_abl_scanset_w0}. This may stem from the fact that \textit{PCN PRE} is already capable of reducing the real scanning error; hence, it likely provides a better starting point for the loss function minimisation during supervised learning, leading to a better outcome than starting from scratch. Furthermore, it can be observed that not all the introduced modifications in \textit{PCNm} lead to improved result. Indeed, additionally feeding a global ear subsample to the DNN results in inferior denoising performance and longer processing time compared to using only a local point patch as input. This suggests that including sparse points sampled from the full ear shape could be detrimental, as they might add irrelevant information for the denoising task while increasing the parameters to be optimised during training. Conversely, providing a limited amount of real scanning data to \textit{PCNm} is beneficial, since the model appears capable of learning the inherent characteristics of the error affecting these point clouds, which shows only a moderate degree of correlation with the synthetic error. The iterative denoising of the \textit{KU1} point clouds using \textit{PCN REF} and \textit{PCNm RET} reduces the overall deviation from photogrammetric to underlying geometry, as evidenced by the \textbf{MD} metric; however, it is not effective in improving \textbf{CD} and \textbf{HD}. These latter two metrics also include the distance between reference and noisy points; thus, they are largely affected by the formation of holes in the denoised point clouds, as those visible in Fig.~\ref{fig:ku1_phm2_scn_pcn_ret_iter}. The fact that \textit{PCNm RET} results in a lower median \textbf{NR} in \textbf{MD}, and a steeper deterioration in \textbf{CD} and \textbf{HD} compared to \textit{PCN REF} with subsequent iterations, may be attributed to the differences in the loss functions between the models. Specifically, for \textit{PCNm RET}, this is designed to better target the concave pinna structure, where higher errors and incompleteness in the scan are often present due to occlusion. The larger deviations and more complex geometry in these locations make them challenging to tackle, leading to poorer outcomes in the employed metrics, which focus on the overall denoising performance across the entire ear geometry.

The analyses conducted in Sec.~\ref{sec:3B} indicate that only a limited set of denoising models is capable of effectively reducing the deviation of the HRTFs computed on the processed geometries compared to those obtained on the initial photogrammetric scan. While \textit{PCNm RET I1} leads to the best outcome, \textit{DMR REF} and \textit{PCN REF I1} result in lower improvements, and all other models yield worse HRTFs than \textit{PHO}. Although a clear explanation for this behaviour is challenging to ascertain, it can be attributed to the specific denoising approaches employed by the various methods, which may lead to oversmoothing or incorrect point displacement in areas close to the concave pinna structures, mainly affected by errors and incompleteness. Promising results are achieved with \textit{PCNm RET}, for which the original \textit{PCN} loss function is modified to specifically target the most occluded pinna locations, and the model is trained from scratch on the synthetic scanned ear dataset and fine-tuned on a limited number of real scans. However, the observed improvements remain marginal. Additionally, iterative denoising with this model and \textit{PCN REF} adversely impacts the HRTFs, likely due to the degradation of the meshed ear geometries after multiple denoising passes. Given that exposing \textit{PCNm} to a small number of samples with real scanning error enhances its denoising performance, using larger datasets of matching photogrammetry and reference individual head and ear geometries, along with a more relevant loss function, is deemed to lead to greater improvements in HRTF accuracy.

The analysis of the HRTFs related to the processed ear point clouds further indicates that the denoising performance, measured through the employed geometric metrics, is not fully correlated with its effects on the HRTFs. Indeed, while \textit{PCNm RET I1} leads to worse denoising performance compared to \textit{PCN REF I1}, the HRTFs assessment metrics exhibit the opposite trend. This discrepancy is attributed to the fact that certain parts of the ear geometry are more sensitive to error than others, and the geometric metrics do not account for it. This is also evident in the outcome of the correlation analysis between HRTFs and geometric metrics, reported in Sec.~\ref{sec:3C}. Focusing on the results obtained when considering the full ear point cloud, it can be observed that among the unweighted \textbf{CD}, \textbf{HD}, and \textbf{MD}, used to evaluate the denoising performance in Sec.~\ref{sec:3A}, only the latter shows a correlation coefficient exceeding \SI{0.25}{}. Nonetheless, using a weight proportional to $\mathrm{AO}^c$ increases the correlation between geometric and HRTF metrics. Therefore, a proper choice of the geometric metric can lead to a better assessment of the scanned ear geometries, especially when the aim is to acquire individual HRTFs. These findings can also guide the selection of a relevant loss function for DNN denoising, such as the weighted \textbf{Acc}, \textbf{Avg}, or \textbf{MD} computed on the full ear, or \textbf{CD} on the pinna front. While focusing on the ear front might be effective, algorithmically extracting frontal points from the ear points clouds remains challenging. A potential solution could involve employing a DNN aimed at shape classification or semantic segmentation of input point clouds to automate the selection of frontal ear points \cite{Qi2017PointNet:Segmentation}.

While analysing the effect of repeated photogrammetric scans is not a primary goal of this work, it is important to note that such variability can be present and may strongly impact the related HRTFs. The analyses carried out on the original and repeated scans conducted on the \textit{KU1} with different optical treatments, although using similar scanning equipment and conditions, seem to indicate that the results may vary depending on several factors, including the expertise of the operator, as also suggested in \cite{Reichinger2013EvaluationPinnas}. In Tab.~\ref{tab:scan_metrics}, while some of the geometric metrics computed on the full ear are similar between the original and repeated scans, e.g.\ \textbf{Cmp}, \textbf{Max}, and \textbf{HD}, others indicate a better outcome for the latter. Attention to properly scanning the most acoustically relevant parts of the ear can greatly benefit the results. This can be achieved by increasing the number of images focusing on the concave pinna structures and the amount of angles from which these are acquired. Indeed, comparing the deviation in localisation performance obtained with the HRTFs computed on the original and repeated photogrammetric scans in Fig.~\ref{fig:stats_qe_pe}, it can be seen that the former relates to an outcome tending towards that of non-individual filters, while the latter is closer to that of correctly acquired individual data. Although the procedure applied for the repeated scans in the current study shows that the related HRTFs are close to reference ones, this scanning is conducted on an optically treated dummy head. The best results for the scanning of $3$ human subjects' ears, using a professional camera setup and spraying the pinnae with black water colour to create a dense random pattern, show averaged values of $\mathrm{\textbf{Acc}} = \SI{1.70}{\milli\meter}$ and $\mathrm{\textbf{Cmp}} = \SI{80.9}{\percent}$ at the front \cite{Reichinger2013EvaluationPinnas}. The reported deviation is higher than that of the original \textit{KU1} scans; hence, it is thought to relate to large errors in modelled localisation performance with the computed HRTFs. This suggests that acquiring the ear geometry of human subjects may pose greater challenges than scanning dummy heads or plaster casts. Therefore, further research is needed to define the optimal procedures and parameters for accurate photogrammetric scanning of human ears. This investigation should evaluate whether direct scanning provides sufficient precision or if additional processing of the raw scans, e.g.\ the denoising approach proposed in the current study, or alternative scanning methods, either as substitutes or complements to photogrammetry, are required.

Given that, in recent years, the task of point cloud denoising has received considerable attention, with several classical and DL-based approaches being developed for this purpose \cite{Zhou2022PointApproaches}, novel denoising techniques could be tested to reduce the scanning error in photogrammetric ear point clouds. For instance, DNN architectures trained to learn implicit surface representations from local and global point cloud features, from which watertight meshes can be directly extracted using algorithms such as marching cubes, have been proposed \cite{Erler2020Points2SurfClouds}. These approaches could avoid the need for meshing algorithms, e.g.\ screened Poisson surface reconstruction used in the current study, requiring parameter tuning to obtain optimal results \cite{Kazhdan2013ScreenedReconstruction}, thereby simplifying the processing steps. Additionally, these methods encode latent descriptions of the underlying global surface from input point clouds, making them more robust to noisy or sparsely sampled inputs, as they obtain a strong prior from the training data \cite{Erler2020Points2SurfClouds}. Consequently, such DNNs are deemed effective on photogrammetric ear scans, as the pinnae, despite individual morphological variations, present a common general shape. Furthermore, the scanned geometries often exhibit incompleteness and large deviations at the most concave pinna structures, whose accurate reconstruction is critical for the HRTF spectral features. However, it remains to be tested whether these approaches can handle the complex shape of scanned ears and the specific errors affecting them, as well as whether they can generate meshes with sufficiently high resolution for numerical HRTF computation.

An alternative method to obtain a reliable mesh from individual ear data is a Parametric Pinna Model (PPM), which deforms an accurate template pinna mesh to match a target ear geometry. In \cite{Pollack2022ParametricGeometry}, a PPM was evaluated by manually aligning its $114$ parameters to match $6$ target pinna meshes, until reaching a mean geometric error below \SI{1}{\milli\meter}. The sagittal plane localisation model was used to compare the HRTFs of the target meshes with those of the deformed template, resulting in maximum deviations in QE and PE of around \SI{4}{\percent} and \SI{3}{\degree}, respectively. These results are slightly worse than those obtained in the current analysis after denoising the repeated \textit{KU1} scans with one iteration of \textit{PCNm RET}. As demonstrated in \cite{Pausch2023ComparisonModel}, DL algorithms can be used to predict a subset of PPM parameters from multi-view-plus-depth images of pinnae. The two DNNs applied for this scope, assessed with a geometric metric akin to \textbf{HD} calculated between deformed template and target meshes, yielded promising results, with an average of \SIlist{0.93; 0.5}{\milli\meter} for the worst and best-performing architectures, respectively. However, as indicated by the correlation study in Sec.~\ref{sec:3C}, the unweighted \textbf{HD} on the full ear shows weak correlation with the HRTF metrics. The current findings suggest that other measures, e.g.\ \textbf{Acc} or \textbf{Avg}, may provide a better quantitative assessment of the match between two pinna geometries. These insights can also aid in selecting an appropriate loss function for the unsupervised training of a DL model leveraging the PPM to synthesise pinna geometries from input point clouds \cite{Perfler2023ConceptsModel}. This would allow training without labelled PPM parameters, enabling the use of pinna point clouds as input data, e.g.\ those already available in several databases or potentially obtainable through photogrammetry. Furthermore, efficient scanning techniques to capture accurate ear geometries for HRTF computation on a large scale, e.g.\ photogrammetry followed by denoising or PPMs, could lead to the creation of extensive datasets for training DL algorithms designed to directly synthesise individual HRTFs from point clouds or images, further simplifying HRTF acquisition. This is deemed beneficial, as these approaches currently often suffer from low generalisability and a lack of accuracy in the reconstructed HRTF monaural spectral cues \cite{Miccini2020HRTFLearning}.

Finally, while the employed sagittal localisation model can provide an indication of the perceptual fit between two individual HRTFs, it only approximates the complex auditory processes involved in human sound localisation \cite{Baumgartner2014ModelingListeners}. Although its outcome is reported to transfer well to stationary localisation error in a VR scene \cite{Jenny2020UsabilityLocalization}, further studies are necessary to perceptually assess the individual HRTFs obtained on the photogrammetric and denoised geometries in realistic VR environments, potentially including multiple sound sources and room acoustic modelling.

\section{Conclusion}
\label{sec:5}
The photogrammetric scans of individual ears for numerical HRTF computation often exhibit large geometric deviations at the most concave pinna structures. While interaural features evaluated on HRTFs computed on photogrammetric dummy head scans show a relatively good match with reference data, monaural features are generally incorrectly estimated, resulting in substantial perceptual deviation from reference HRTFs, as assessed by a sagittal plane localisation model.

To address these issues, the use of DL algorithms designed for point cloud denoising is proposed. Three distinct DNN architectures are tested and benchmarked against a classical denoising approach. To enhance their performance on photogrammetric ear scans, the DNNs are fine-tuned on a dataset of pinna geometries corrupted with synthetic error, mimicking that observed in the available photogrammetric scans of optically treated dummy head ears, comprising a limited set of $12$ samples. Modifications to the loss function and architecture of one DNN are applied, aiming to target the denoising on the most concave pinna locations, where the geometric deviation is critical. This model is retrained from scratch on the synthetic ear dataset and further fine-tuned on a limited set of scanned samples to evaluate the impact of exposing it to real scanning error, given the moderate correlation between this and the synthetic error. Various geometric metrics are used to assess the performance of the denoising models on datasets containing ear shapes affected by synthetic and real scanning error. 

An efficient FEM formulation is used to compute the HRTFs on meshes derived from the scanned point clouds, as its results exhibit a better match with the analytical solution of a simplified case compared to those of a BEM approach typically used for such computations. Objective and perceptually inspired metrics are employed to compare HRTFs derived from initial photogrammetric and denoised geometries, and those from accurate reference scans.

Three geometric metrics evaluated on the full ear point clouds are employed to assess the performance of the denoising models. The results of the original versions of the tested DNNs are generally in line with or worse than those of a classical denoising approach. Although fine-tuning the DL models on the synthetic ear dataset, and testing them on similarly corrupted ear shapes, yields benefits for some architectures, only one DNN shows improved denoising of the real scanning error, outperforming the classical approach. The modifications applied to the best performing DNN do not consistently improve its outcome; indeed, the model with modified architecture and loss function yields worse results than the version including only the loss function modification. Moreover, fine-tuning this model on a limited set of samples containing real scanning error further enhances denoising performance, but this tends to be inferior to that of the original DNN version refined on the ear dataset. While both the original and modified models can be applied iteratively to reduce residual error, multiple iterations tend to introduce holes in the point clouds, leading to artefacts in the meshes generated from these geometries.

The assessment of the HRTFs computed on the denoised geometries indicates that only certain DNNs lead to better results than those of the raw photogrammetric data. The best results are achieved with the modified DNN, retrained on synthetic ear samples and fine-tuned on a subset of real scans. The sagittal localisation analysis conducted on the HRTFs of the denoised ear scans with this model shows a median deviation close to that of accurately acquired individual filters; however, the achieved enhancement is marginal and can likely be improved with the choice of a more relevant loss function.

The geometric error analysis of original and repeated photogrammetric KU100 dummy head scans suggests that the results depend on the expertise of the operator. While the HRTFs from the original scans exhibit large deviations in modelled perceptual metrics, analogous to those of non-individual filters, the repeated scan HRTFs align more closely with correctly measured individual data.

Correlation analysis of the geometric metrics evaluated on the full ear point clouds and those computed on their HRTFs reveals coefficients below \SI{0.5}{}. Weighting the distance by a factor proportional to the complement of the ambient occlusion improves correlation, emphasising acoustically relevant ear locations. Additionally, the correlation can be further increased by focusing solely on deviations in the frontal part of the pinna. The most correlated geometric metrics for the full ear are Accuracy, Average, and Mesh Distance, while Chamfer Distance shows the best performance when considering only the pinna front. The insights gained from identifying the most correlated geometric metrics between a target and reference ear geometry, in terms of similarity of their related HRTFs, can guide the selection of optimal loss functions for training DNNs aimed at directly or indirectly acquiring accurate personalised HRTFs from individual input data.

These findings emphasise the need for further research to determine whether photogrammetry can effectively capture individual ear geometries, or if additional processing, such as the proposed denoising approach or alternative techniques, is necessary. Additionally, since the outcome is currently based on modelled sound localisation results, experiments in complex VR environments are essential to validate the perceptual similarity of the acquired individual HRTFs.

\section*{CRediT author statement}
\textbf{Fabio Di Giusto:} Conceptualization, Methodology, Software, Validation, Formal analysis, Investigation, Data Curation, Writing - Original Draft, Visualization.

\textbf{Francesc Lluís:} Software, Investigation, Resources, Writing - Review \& Editing.

\textbf{Sjoerd van Ophem:} Writing - Review \& Editing, Supervision, Funding acquisition.

\textbf{Elke Deckers:} Writing - Review \& Editing, Supervision, Funding acquisition.

\section*{Conflict of interest}
The authors declare no conflict of interest.

\section*{Data availability statement}
Part of the research data and code are available in the KU Leuven Research Data Repository (RDR) and can be found at \url{https://doi.org/10.48804/JYELX1}.
Additional data will be made available on request.

\section*{Acknowledgements}
The European Commission is gratefully acknowledged for their support of the VRACE research project (GA 812719). Internal Funds KU Leuven are gratefully acknowledged for their support. The resources and services used in this work were provided by the VSC (Flemish Supercomputer Center), funded by the Research Foundation - Flanders (FWO) and the Flemish Government.



\small
\begin{thebibliography}{10}

\bibitem{Katz2018BinauralReproduction}
B.~F. Katz and R.~Nicol, ``{Binaural Spatial Reproduction},'' in {\em Sensory Evaluation of Sound} (N.~Zacharov, ed.), ch.~11, pp.~349--388, CRC Press, Taylor {\&} Francis Group, 1st editio~ed., 2018.

\bibitem{Baumgartner2014ModelingListeners}
R.~Baumgartner, P.~Majdak, and B.~Laback, ``{Modeling sound-source localization in sagittal planes for human listeners},'' {\em The Journal of the Acoustical Society of America}, vol.~136, pp.~791--802, 8 2014.

\bibitem{Pollack2022PerspectiveOverview}
K.~Pollack, W.~Kreuzer, and P.~Majdak, ``{Perspective Chapter: Modern Acquisition of Personalised Head-Related Transfer Functions - An Overview},'' in {\em Advances in Fundamental and Applied Research on Spatial Audio}, pp.~1--36, IntechOpen, 4 2022.

\bibitem{Jenny2020UsabilityLocalization}
C.~Jenny and C.~Reuter, ``{Usability of Individualized Head-Related Transfer Functions in Virtual Reality: Empirical Study With Perceptual Attributes in Sagittal Plane Sound Localization},'' {\em JMIR Serious Games}, vol.~8, no.~3, pp.~1--15, 2020.

\bibitem{Zhong2014Head-RelatedDisplay}
X.-l. Zhong and B.-s. Xie, ``{Head-Related Transfer Functions and Virtual Auditory Display},'' in {\em Soundscape Semiotics - Localisation and Categorisation} (H.~Glotin, ed.), vol.~1, ch.~6, pp.~99--134, IntechOpen, 2014.

\bibitem{Ziegelwanger2013CalculationQuality}
H.~Ziegelwanger, A.~Reichinger, and P.~Majdak, ``{Calculation of listener-specific head-related transfer functions: Effect of mesh quality},'' in {\em Proceedings of Meetings on Acoustics}, vol.~19, (Montreal, Canada), pp.~1--8, Acoustical Society of America (ASA), 2013.

\bibitem{DiGiusto2023AnalysisHead}
F.~Di~Giusto, S.~van Ophem, W.~Desmet, and E.~Deckers, ``{Analysis of laser scanning and photogrammetric scanning accuracy on the numerical determination of Head-Related Transfer Functions of a dummy head},'' {\em Acta Acustica}, vol.~7, pp.~1--21, 10 2023.

\bibitem{Reichinger2013EvaluationPinnas}
A.~Reichinger, P.~Majdak, R.~Sablatnig, and S.~Maierhofer, ``{Evaluation of Methods for Optical 3-D Scanning of Human Pinnas},'' in {\em Proceedings - 2013 International Conference on 3D Vision}, (Seattle, USA), pp.~390--397, IEEE, 2013.

\bibitem{Luo2021Score-BasedDenoising}
S.~Luo and W.~Hu, ``{Score-Based Point Cloud Denoising},'' in {\em Proceedings of the IEEE/CVF International Conference on Computer Vision}, pp.~4583--4592, 2021.

\bibitem{Rakotosaona2019PointCleanNet:Clouds}
M.-J. Rakotosaona, V.~La~Barbera, P.~Guerrero, N.~J. Mitra, and M.~Ovsjanikov, ``{PointCleanNet: Learning to Denoise and Remove Outliers from Dense Point Clouds},'' {\em Computer Graphics Forum}, vol.~39, no.~1, pp.~185--203, 2019.

\bibitem{Miccini2020HRTFLearning}
R.~Miccini and S.~Spagnol, ``{HRTF Individualization using Deep Learning},'' in {\em Proceedings of 2020 IEEE Conference on Virtual Reality and 3D User Interfaces Abstracts and Workshops (VRW)}, (Atlanta, GA), pp.~390--395, 2020.

\bibitem{Andreopoulou2015Inter-LaboratoryComparison}
A.~Andreopoulou, D.~R. Begault, and B.~F. Katz, ``{Inter-Laboratory Round Robin HRTF Measurement Comparison},'' {\em IEEE Journal on Selected Topics in Signal Processing}, vol.~9, no.~5, pp.~895--906, 2015.

\bibitem{Dinakaran2018PerceptuallySystems}
M.~Dinakaran, F.~Brinkmann, S.~Harder, R.~Pelzer, P.~Grosche, R.~R. Paulsen, and S.~Weinzierl, ``{Perceptually Motivated Analysis of Numerically Simulated Head-Related Transfer Functions Generated by Various 3D Surface Scanning Systems},'' in {\em ICASSP, IEEE International Conference on Acoustics, Speech and Signal Processing - Proceedings}, pp.~551--555, IEEE, 2018.

\bibitem{Cignoni1998Metro:Surfaces}
P.~Cignoni, C.~Rocchini, and R.~Scopigno, ``{Metro: measuring error on simplified surfaces},'' {\em Computer Graphics Forum}, vol.~17, no.~2, pp.~167--174, 1998.

\bibitem{Cignoni2008MeshLab:Tool}
P.~Cignoni, M.~Callieri, M.~Corsini, M.~Dellepiane, F.~Ganovelli, and G.~Ranzuglia, ``{MeshLab: an Open-Source Mesh Processing Tool},'' in {\em Sixth Eurographics Italian Chapter Conference}, pp.~129--136, The Eurographics Association, 2008.

\bibitem{Rychtarikova2009BinauralRooms}
M.~Rycht{\'{a}}rikov{\'{a}}, T.~Van~den Bogaert, J.~Wouters, and G.~Vermeir, ``{Binaural sound source localization in real and virtual rooms},'' {\em AES: Journal of the Audio Engineering Society}, vol.~57, no.~4, pp.~205--220, 2009.

\bibitem{Rychtarikova2011PerceptualUnderstanding}
M.~Rycht{\'{a}}rikov{\'{a}}, T.~van~den Bogaert, G.~Vermeir, and J.~Wouters, ``{Perceptual validation of virtual room acoustics: Sound localisation and speech understanding},'' {\em Applied Acoustics}, vol.~72, no.~4, pp.~196--204, 2011.

\bibitem{Sinev2023IndividualPinnae}
D.~Sinev, F.~Di~Giusto, J.~Peissig, S.~van Ophem, and E.~Deckers, ``{Individual Ear Replicas with Complete Ear Canals Compatible with an Artificial Head Pinnae},'' in {\em Proceedings of DAGA 2023}, (Hamburg, Germany), pp.~1--4, 2023.

\bibitem{Roden2020TheResearch}
R.~Roden and M.~Blau, ``{The IHA database of human geometries including torso, head and complete outer ears for acoustic research},'' in {\em Proceedings of 2020 International Congress on Noise Control Engineering, INTER-NOISE 2020}, (Seoul, Korea), pp.~4226--4237, 2020.

\bibitem{Schonberger2016Structure-from-MotionRevisited}
J.~L. Sch{\"{o}}nberger and J.-M. Frahm, ``{Structure-from-Motion Revisited},'' in {\em Conference on Computer Vision and Pattern Recognition (CVPR)}, (Las Vegas, NV, USA), pp.~4104--4113, IEEE, 2016.

\bibitem{Schonberger2016PixelwiseStereo}
J.~L. Sch{\"{o}}nberger, E.~Zheng, M.~Pollefeys, and J.~M. Frahm, ``{Pixelwise View Selection for Unstructured Multi-View Stereo},'' in {\em European Conference on Computer Vision (ECCV)}, (Amsterdam, The Netherlands), pp.~1--15, Springer, 2016.

\bibitem{Taskesen_distfit_is_a_2020}
E.~Taskesen, ``{distfit is a python library for probability density fitting.},'' 1 2020.

\bibitem{Brinkmann2019AResponses}
F.~Brinkmann, M.~Dinakaran, R.~Pelzer, P.~Grosche, D.~Voss, and S.~Weinzierl, ``{A cross-evaluated database of measured and simulated HRTFs including 3d head meshes, anthropometric features, and headphone impulse responses},'' {\em AES: Journal of the Audio Engineering Society}, vol.~67, no.~9, pp.~705--718, 2019.

\bibitem{Jacobson2018libigl:Library}
A.~Jacobson, D.~Panozzo, and {Others}, ``{libigl: A simple C++ geometry processing library},'' 2018.

\bibitem{Luo2020DifferentiableDenoising}
S.~Luo and W.~Hu, ``{Differentiable Manifold Reconstruction for Point Cloud Denoising},'' in {\em Proceedings of the 28th ACM International Conference on Multimedia}, pp.~1330--1338, 2020.

\bibitem{Guerrero2018PCPNet:Clouds}
P.~Guerrero, Y.~Kleiman, M.~Ovsjanikov, and N.~J. Mitra, ``{PCPNet: Learning Local Shape Properties from Raw Point Clouds},'' {\em Computer Graphics Forum}, vol.~37, no.~2, p.~75–85, 2018.

\bibitem{Egner2023PolynomialPlates}
F.~S. Egner, L.~Sangiuliano, R.~F. Boukadia, S.~van Ophem, W.~Desmet, and E.~Deckers, ``{Polynomial filters for camera-based structural intensity analysis on curved plates},'' {\em Mechanical Systems and Signal Processing}, vol.~193, p.~110245, 6 2023.

\bibitem{Stitt2021SensitivityModel}
P.~Stitt and B.~F.~G. Katz, ``{Sensitivity analysis of pinna morphology on head-related transfer functions simulated via a parametric pinna model},'' {\em The Journal of the Acoustical Society of America}, vol.~149, no.~4, pp.~2559--2572, 2021.

\bibitem{Ghorbal2017PinnaSets}
S.~Ghorbal, T.~Auclair, C.~Soladi{\'{e}}, and R.~S{\'{e}}guier, ``{Pinna morphological parameters influencing HRTF sets},'' in {\em Proceedings of the 20th International Conference on Digital Audio Effects (DAFx-17)}, (Edinburgh, UK), pp.~353--359, 2017.

\bibitem{Erler2020Points2SurfClouds}
P.~Erler, P.~Guerrero, S.~Ohrhallinger, N.~J. Mitra, and M.~Wimmer, ``{Points2Surf Learning Implicit Surfaces from Point Clouds},'' in {\em Lecture Notes in Computer Science}, pp.~108--124, Springer International Publishing, 2020.

\bibitem{Muntoni2021PyMeshLab}
A.~Muntoni and P.~Cignoni, ``{PyMeshLab},'' 2021.

\bibitem{Mobius2010OpenFlipper:Framework}
J.~M{\"{o}}bius and L.~Kobbelt, ``{OpenFlipper: An Open Source Geometry Processing and Rendering Framework},'' in {\em Proceedings of the 7th international conference on Curves and Surfaces}, (Avignon, France), pp.~488--500, Springer, 7 2010.

\bibitem{Ziegelwanger2016AFunctions}
H.~Ziegelwanger, W.~Kreuzer, and P.~Majdak, ``{A priori mesh grading for the numerical calculation of the head-related transfer functions},'' {\em Applied Acoustics}, vol.~114, pp.~99--110, 2016.

\bibitem{Beriot2016EfficientProblems}
H.~B{\'{e}}riot, A.~Prinn, and G.~Gabard, ``{Efficient implementation of high-order finite elements for Helmholtz problems},'' {\em International Journal for Numerical Methods in Engineering}, vol.~106, pp.~213--240, 2016.

\bibitem{Beriot2020AnShape}
H.~B{\'{e}}riot and A.~Modave, ``{An automatic perfectly matched layer for acoustic finite element simulations in convex domains of general shape},'' {\em International Journal for Numerical Methods in Engineering}, vol.~122, no.~5, pp.~1239--1261, 2020.

\bibitem{DiGiusto2023AnalysisReplicas}
F.~Di~Giusto, D.~Sinev, K.~Pollack, S.~van Ophem, and E.~Deckers, ``{Analysis of Impedance Effects on Head-Related Transfer Functions of 3D Printed Pinna and Ear Canal Replicas},'' in {\em Proceedings of the 10th Convention of the European Acoustics Association Forum Acusticum 2023}, (Turin, Italy), pp.~323--330, European Acoustics Association, 9 2023.

\bibitem{Jin2014CreatingDatabase}
C.~T. Jin, P.~Guillon, N.~Epain, R.~Zolfaghari, A.~Van~Schaik, A.~I. Tew, C.~Hetherington, and J.~Thorpe, ``{Creating the Sydney York Morphological and Acoustic Recordings of Ears Database},'' {\em IEEE Transactions on Multimedia}, vol.~16, no.~1, pp.~37--46, 2014.

\bibitem{Brinkmann2017AOrientations}
F.~Brinkmann, A.~Lindau, S.~Weinzierl, S.~Van De~Par, M.~M{\"{u}}ller-Trapet, R.~Opdam, and M.~Vorl{\"{a}}nder, ``{A High Resolution and Full-Spherical Head-Related Transfer Function Database for Different Head-Above-Torso Orientations},'' {\em Journal of the Audio Engineering Society}, vol.~65, no.~10, pp.~841--848, 2017.

\bibitem{Jacobsen2013FundamentalsAcoustics}
F.~Jacobsen and P.~Moller~Juhl, {\em {Fundamentals of General Linear Acoustics}}.
\newblock Wiley, 1st~ed., 2013.

\bibitem{Pollack2023SpectralEardrum}
K.~Pollack, F.~Di~Giusto, D.~Sinev, and P.~Majdak, ``{Spectral and psychoacoustic evaluation of head-related transfer functions calculated at the blocked ear canal and the eardrum},'' in {\em Proceedings of the 10th Convention of the European Acoustics Association Forum Acusticum 2023}, (Turin, Italy), pp.~1159--1164, European Acoustics Association, 9 2023.

\bibitem{Denk2018SpectralEars}
F.~Denk, S.~D. Ewert, and B.~Kollmeier, ``{Spectral directional cues captured by hearing device microphones in individual human ears},'' {\em The Journal of the Acoustical Society of America}, vol.~144, no.~4, pp.~2072--2087, 2018.

\bibitem{Majdak2022AMTModeling}
P.~Majdak, C.~Hollomey, and R.~Baumgartner, ``{AMT 1.x: A toolbox for reproducible research in auditory modeling},'' {\em Acta Acustica}, vol.~6, no.~19, pp.~1--17, 2022.

\bibitem{Qi2017PointNet:Segmentation}
C.~R. Qi, H.~Su, K.~Mo, and L.~J. Guibas, ``{PointNet: Deep learning on point sets for 3D classification and segmentation},'' in {\em Proceedings - 30th IEEE Conference on Computer Vision and Pattern Recognition, CVPR 2017}, pp.~1--19, 2017.

\bibitem{Zhou2022PointApproaches}
L.~Zhou, G.~Sun, Y.~Li, W.~Li, and Z.~Su, ``{Point cloud denoising review: from classical to deep learning-based approaches},'' {\em Graphical Models}, vol.~121, no.~101140, pp.~1--16, 2022.

\bibitem{Kazhdan2013ScreenedReconstruction}
M.~Kazhdan and H.~Hoppe, ``{Screened Poisson Surface Reconstruction},'' {\em ACM Transactions on Graphics}, vol.~32, no.~3, pp.~1--13, 2013.

\bibitem{Pollack2022ParametricGeometry}
K.~Pollack, F.~Pausch, and P.~Majdak, ``{Parametric pinna model for a realistic representation of listener-specific pinna geometry},'' in {\em Proceedings of the 24th International Congress on Acoustics}, (Gyeongju, Korea), pp.~168--178, 2022.

\bibitem{Pausch2023ComparisonModel}
F.~Pausch, F.~Perfler, N.~Holighaus, and P.~Majdak, ``{Comparison of deep-neural-network architectures for the prediction of head-related transfer functions using a parametric pinna model},'' in {\em Proceedings of the 10th Convention of the European Acoustics Association Forum Acusticum 2023}, (Turin, Italy), pp.~2329--2334, European Acoustics Association, 9 2023.

\bibitem{Perfler2023ConceptsModel}
F.~Perfler, P.~Majdak, F.~Pausch, and K.~Pollack, ``{Concepts for the evaluation of a parametric pinna model},'' in {\em Proceedings of the 10th Convention of the European Acoustics Association Forum Acusticum 2023}, (Turin, Italy), pp.~2335--2338, European Acoustics Association, 9 2023.

\end{thebibliography}
\end{document}